\documentstyle[12pt]{article}
\author{A. A. Arkhipov\\
\it Institute for High Energy Physics, Protvino,\\
\it Moscow Region, 142284 Russia}
\title{{\bf Quark--Quark  Forces in Quantum Chromodynamics}\thanks {This article is the contribution to the XXVIIth International Conference on High Energy Physics "ICHEP94" held at 20-27 July 1994, Glasgow, Scotland, UK. There is a print of this manuscript, as itself, in the library of the Glasgow University only, and the reference in Proceedings of the "ICHEP94", IOP Publishing Ltd, 1995, Volume II: Parallel Sessions, p. 1430, available at the CERN server as well there is.}}
\date{}
\begin{document}
\vspace{-2cm}
\maketitle
\vspace{-1cm}
\begin{abstract}
By single-time reduction technique of Bethe-Salpeter formalism for
two-fermion systems analytical expressions for the quasipotential of
quark-quark interactions in QCD have been obtained in one-gluon
exchange approximation. The influence of infrared singularities of
gluon Green`s functions on the character of quark-quark forces in QCD
has been investigated. The way the asymptotic freedom manifests itself
in terms of two-quark interaction quasipotential in quantum
chromodynamics is shown. Consistent relativistic consideration of quark
interaction problem by single-time reduction technique in QFT allows
one to establish a nontrivial energy dependence of the two-quark
interaction quasipotential. As a result of the energy dependence of
the interaction quasipotential, the character of the forces changes
qualitatively during the transition from the discrete spectrum
(the region of the negative values of the binding energy) to the
continuous spectrum (that of the positive values of the binding
energy): the smooth behaviour of the interaction quasipotential in
the discrete spectrum goes into the oscillation in the continuous
spectrum. This result gives a visual physical picture where the
oscillations may be interpreted as a manifestation of a quasicrystall
structure of the vacuum.
\end{abstract}
\def\ee{\end{equation}}
\def\be{\begin{equation}}
\def\dint{\mathop{\intop\kern-0.5em\intop}}
\def\ber{\begin{eqnarray}}
\def\eer{\end{eqnarray}}
\def\ovc#1{\displaystyle\mathop{#1}^{\kern0.2em\circ}}

\section{Introduction}

The investigation of relativistic systems interaction properties is
one of the central point in elementary particle physics and the physics
of atomic nucleus. Two-fermion systems are the very tool with the help
of which we study and check our notions about fundamental forces acting
in Nature. The hydrogen atom is the most well-known example of a two-fermion
system. Figuratively speaking one may say that the hydrogen atom was the
laboratory where has been created, checked and studied one of the wonderful
physical theory of the XX century, i.e., quantum mechanics. The first checks
of quantum electrodynamics were also carried out on the two-fermion systems
such as hydrogen atom, positronium, muonium, etc. [1]. The study of deuteron
as a system consisting of two nucleons allowed one to go considerably forth
in our understanding of nuclear forces [2]. The problem of nucleon-nucleon
interaction is still one of the fundamental problems in nuclear physics.

The observation of the $J/\Psi$ and $\Upsilon$ mesons consisting of
$c\bar c$ and $b\bar b$ quarks stimulated our interest in a theoretical
description of the properties of two-fermion systems. At present in the
framework of the local quantum field theory we have a fundamental gauge
model for strong interactions known as quantum chromodynamics (QCD).
The problem of describing the spectroscopy of hadronic states, as bound
states of quark and gluon fields, remains to be one of the important but
unsolved problems of QCD.

At the same time the spectroscopy of the $J/\Psi$ and $\Upsilon$
particle families finds its excellent description in the framework of
the potential models with phenomenological potentials [3,4]. In this,
a very interesting fact has been revealed, the spectroscopy of quarkonium
systems has many features similar to those of the spectroscopy of
hydrogen atom and positronium. This is the reason why the quarkonium
system where the interaction between quarks is given by Coulomb-like
potential, added with a term linearly growing with distance, is
sometimes declared to be a "hydrogen atom" for strong-interacting
systems.

The success of the potential models in describing the spectroscopy
of quarkonium systems may be thought to be not accidental and in order
to understand why it is so, one should, first of all, clarify what is the
status of the potential models in the framework of QCD. The solution of
this problem would allow one to establish the connection between the
fundamental theory and experiment right at the point where at present
we have confrontation. Note that indicated point of the confrontation
between theory and experiment is an essentially nonperturbative region.

In a more general case this problem may be formulated as a problem
of calculating the interaction potential between quarks proceeding
from the first principles. In the case the problem be solved one
would manage to predict all the properties of quark systems proceeding
from the fundamental QCD Lagrangian. Here great hopes are set on the
calculations based on the lattice methods and large efforts are
undertaken in this direction [5-7].

In the present report we are going to show that there exist a simpler
and more consistent way to solve this problem, which is based on the
single-time formalism in quantum field theory (QFT). Here we will present
the results of our last works where the problem of calculating the
quark-quark forces in QCD has been considered. We have used the
single-time formalism in QFT as a tool in the investigation of the
problem.

\section{Necessary Information on Single--Time \newline Reduction
Technique}

As usually we shall introduce the Bethe--Salpeter wave function
of the composite two--fermion quark--antiquark system with the
help of the matrix element
\[
\Phi _i^j(x_1x_2) = <0\mid T(\Psi _i(x_1)\bar\Psi ^j(x_2))\mid \Phi >,
\]
where $\Psi _i(x_i)$ are Heisenberg operators of the quark
fields, $\mid \Phi >$ is
a normalized vector of the bound state of quark--antiquark
system. Heisenberg operators have an index, which in the case of
QCD is complicated and is a set of three indices:
$i=(\alpha _1,f_1,c_1)$, $j=(\alpha _2,f_2,c_2)$,
where $\alpha $ is the spinor index, $f$ is the flavour index,
and $c$ is the colour index. The single--time wave function of
quark--antiquark system is defined with the equation [8]
\ber
\lefteqn{\tilde \Psi (n\tau \mid x_1x_2) = } \label{2}\\
& & =\frac{1}{i^2}\dint_{n\zeta _1=\tau =n\zeta _2}S_1^{(-)}(x_1+n\tau
-\zeta _1)d\hat\sigma _{\zeta _1}
\Phi (\zeta _1\zeta _2)d\hat\sigma _{\zeta _2}S_2^{(+)}(\zeta _2-n\tau
-x_2). \nonumber
\eer
In the R.H.S. of Eq. (\ref{2}) we have not only integration over
space--like surface but summation over indices, which we do
not write down explicitly. Here one should bear in mind that
\[
[S_1^{(-)}(x)]_i^{i'} = \delta _f^{f'}\delta _c^{c'}[S_1^{(-)}]_\alpha
^{\alpha '},\quad
[S_2^{(+)}]_{j'}^j = \delta _{f'}^f\delta _{c'}^c[S_2^{(+)}]_{\beta
'}^\beta ,
\]
where $[S_1^{(-)}(x)]_\alpha ^{\alpha '}$, \ \ $[S_2^{(+)}(x)]_{\beta
'}^\beta $
are frequency parts of the
one--particle causality Green's function of the spinor fields and
\[
[d\hat\sigma _{\zeta _1}]_i^{i'} = \delta _f^{f'}\delta _c^{c'}[\gamma _\mu
]_\alpha ^{\alpha '}d\sigma _{\zeta _1}^\mu ,
\quad [d\hat\sigma _{\zeta _2}]_j^{j'} = \delta _f^{f'}\delta
_c^{c'}[\gamma _\mu ]_\beta ^{\beta '}
d\sigma _{\zeta _2}^\mu ,
\]
where $d\sigma _\zeta ^\mu $ is a differential element of a flat
space--like
surface at the point $\zeta $. The surface is given by the Eq.
$n\zeta =\tau $, where $n^\mu (n^2=1)$ is a unit time--like vector of the
normal to
the given flat hypersurface.

The momentum representation for single--time wave function
(\ref{2})
is introduced with the following integral transformation [8]
\ber
\lefteqn{\Psi (n\tau \mid \tilde{p}_1\sigma _1\tilde{p}_2\sigma _2)
=}\nonumber\\
& &= \dint\bar{u}(x_1\mid \vec{p}_1\sigma _1)
d\hat\sigma _{x_1}\tilde\Psi (n\tau \mid x_1x_2)d\hat\sigma _{x_2}
v(x_2\mid \vec{p}_2\sigma _2) ,\label{3}
\eer
where in the R.H.S. of Eq. (\ref{3}) the integration is carried out
over some space--like surfaces and one can easily verify that the
result of such integration does not depend on the choice of these
surfaces. Summation over the indices, which
is not explicitly given here, however implied. The index $\sigma $ is a
compound one and is a set of three indices $\sigma =(\sigma _\alpha ,\sigma
_f,\sigma _c)$. The
functions $u$ and $v$ are one--particle wave functions of quark and
antiquark, respectively, which satisfy the normalization
conditions
\ber
\int\bar{u}(x\mid \vec{p}\sigma )d\hat\sigma _{x}u(x\mid \vec{k}\sigma ')&
= &
\int\bar{v}(x\mid \vec{p}\sigma )d\hat\sigma _{x}v(x\mid \vec{k}\sigma ') =
\nonumber\\
& = &2E(\vec{p})\delta ^3(\vec{p}-\vec{k})\delta _{\sigma \sigma
'}\nonumber
\eer
and of completeness
\[
\sum_\sigma \int{d\mu (\vec{p})u(x\mid \vec{p}\sigma )\bar{u}(y\mid
\vec{p}\sigma )}=\frac{1}{i}
S^{(-)}(x-y),
\]
\[
\sum_\sigma \int{d\mu (\vec{p})v(x\mid \vec{p}\sigma )\bar{v}(y\mid
\vec{p}\sigma )}=\frac{1}{i}
S^{(+)}(x-y),
\]
where $d\mu (\vec{p})$ is an invariant measure in the momentum space (an
element of the one--particle phase volume)
\[
d\mu (\vec{p})=(2\sqrt{m^2+\vec{p}^2})^{-1}d^3\vec{p}.
\]
Everywhere tilde above the momentum implies, that the given
momentum lies on the mass shell $\tilde{p}_i^2 = m_i^2$.

Introduce also the Fourier transformation of the single--time
wave function over the variable $\tau $
\[
\Psi (nM\mid \tilde{p}_1\sigma _1\tilde{p}_2\sigma
_2)=\int_{-\infty}^{\infty}{d\tau exp
(iM\tau )\Psi (n\tau \mid \tilde{p}_1\sigma _1\tilde{p}_2\sigma _2)}.
\]
For the single--time wave function thus defined in ref.[8] we
obtained a three--dimensional dynamic equation, which has the
form
\ber
\Psi (nM\mid \tilde{p}_1\sigma _1\tilde{p}_2\sigma
_2)=(n\tilde{p}_1+n\tilde{p}_2-M)^
{-1}\sum_{\lambda _1\lambda _2}\dint d\mu _1(\vec{k}_1)d\mu
(\vec{k}_2)\times \nonumber\\
\times V(nM\mid \tilde{p}_1\sigma _1\tilde{p}_2\sigma _2;\tilde{k}_2\lambda
_2\tilde{k}_1\lambda _1)
\Psi (nM\mid \tilde{k}_1\lambda _1\tilde{k}_2\lambda _2).\label{4}
\eer
The function $V$ in the R.H.S. of Eq. (\ref{4}) describes the
quark--antiquark interaction and is defined with the equation
\ber
V(nM\mid \tilde{p}_1\sigma _1\tilde{p}_2\sigma _2;\tilde{k}_2\lambda
_2\tilde{k
}_1\lambda _1)=
T(nM\mid \tilde{p}_1\sigma _1\tilde{p}_2\sigma _2;\tilde{k}_2\lambda
_2\tilde{k}_1\lambda _1)
\nonumber\\-\sum_{\sigma '_1\sigma '_2}\dint
d\mu (\vec{p'_1})d\mu
(\vec{p'_2})(n\tilde{p'_1}+n\tilde{p'_2}-M)^{-1}\times
\nonumber\\
\times V(nM\mid \tilde{p}_1\sigma _1\tilde{p}_2\sigma _2;\tilde{p'}_2\sigma
'_2\tilde{p'}_1
\sigma '_1)T(nM\mid \tilde{p'}_1\sigma '_1\tilde{p'}_2\sigma
'_2;\tilde{k}_2\lambda _2\tilde
{k}_1\lambda _1).\label{5}
\eer
where the function $T$ is given by the equality
\ber
T(nM\mid \tilde{p}_1\sigma _1\tilde{p}_2\sigma _2;\tilde{k}_2\lambda
_2\tilde{k}_1\lambda _1)=
in_0\delta ^3(\vec{P}_M-\vec{K}_M)\nonumber\\
\bar{u}(\vec{p}_1\sigma _1)v(\vec{p}_2\sigma _2)(2\pi )^{-5}
\int_{-\infty }^{\infty }d\alpha (\frac{1}{\Delta /2+\alpha -i\varepsilon
}+\frac{1}{\Delta /2-\alpha -i\varepsilon })\times
\nonumber\\\times
\int_{-\infty }^{\infty }d\beta (\frac{1}{\Delta '/2+\beta -i\varepsilon
}+\frac{1}{\Delta '/2-\beta -i\varepsilon })\times
\nonumber \\\times \tilde{R}(p-\alpha n;k-\beta n\mid
K_M)u(\vec{k}_1\lambda _1)\bar{v}
(\vec{k}_2\lambda _2),\label{6}
\eer
where $R(p;k\mid K)$ is the Fourier--image of the $VEV$ of the forth
order radiation operator
\ber
\tilde{R}(p_1p_2;k_2k_1)&=&(2\pi )^4\delta ^4(P-K)
\tilde{R}(p;k\mid K),\nonumber\\
\tilde{R}(p_1p_2;k_2k_1)&=&\int dx_1dx_2dy_1dy_2\tilde{R}^{(4)}
(x_1x_2;y_2y_1)\times \nonumber\\&\times &
exp(ip_1x_1+ip_2x_2-ik_1y_1-ik_2y_2),\nonumber\\
\tilde{R}^{(4)}(x_1x_2;y_2y_1)&=&R^{(4)}(x_1y_2;x_2y_1),
\nonumber\\
R^{(4)}(x_1x_2;y_2y_1)&=&
\frac{1}{i^2}<0\mid \frac{\delta ^4S}{\delta \bar\psi (x_1)\delta \bar\psi
(x_2)\delta \psi (y_2)
\delta \psi (y_1)}S^{+}\mid 0>. \nonumber
\eer
In another words the function $R^{(4)}$ is the current
Green's function because it coincides with the VEV of chronological
product of fermion currents up to quasilocal terms ( for details refer to
[8,9]). The following notations are used in the R.H.S. of
Eq.(\ref{6})
\ber
P = \tilde{p}_1+\tilde{p}_2,\quad p = \frac{1}{2}(\tilde{p}_1-
\tilde{p}_2),\quad P_M = P-\Delta n, \nonumber\\
K = \tilde{k}_1+\tilde{k}_2,\quad k = \frac{1}{2}(\tilde{k}_1-
\tilde{k}_2),\quad K_M = K-\Delta 'n, \nonumber\\
\Delta = nP-M,\quad \Delta ' = nK-M. \nonumber
\eer
Besides a convolution over the indices, not written down
explicitly, is implied in the R.H.S of Eq. (\ref{6}).

It is worth stress that the Schr$\ddot o$dinger structure of the
dynamic equation for single-time wave function arises as a consequence
of the causality structure of local quantum field theory. When deriving
the three-dimensional dynamic equations we did not bare in mind any
concrete model of the QFT, but used its most general properties.
Therefore the dynamic equation (\ref{4}) may serve a reliable
foundation for phenomenological investigation of relativistic
two-fermion systems. At the same time the  single--time
reduction technique may serve as effective tool to
investigate any particular quantum field theory model. As such a
particular model, the gauge model was picked  up
known as quantum chromodynamics. On the one hand, this choice
was partially due to that, quantum  chromodynamics  claims  to
describe the hadronic sector of  the  so--called  Standard  Model
and, on the other hand, includes, as a special case, another  gauge
model called quantum electrodynamics (QED), the latter describing
excellently electromagnetic interactions in particle physics.

The calculation of two-quark interaction quasipotential in QCD was
carried out in three stages. The one-gluon approximation has been
used in the first stage. Afterwards the influence of infrared
singularities of gluon Green's functions on the character of
two-quark forces in QCD has been investigated. Finally
the way the asymptotic freedom manifests itself in terms of
two-quark interaction quasipotential in quantum chromodynamics
was shown.

\section{One--Gluon Exchange Approximation \newline in QCD}

Using the QCD Lagrangian structure, we obtain
for  the function  $\tilde  R$  in  the  R.H.S.  of  Eq.   (\ref{6})
in the second order over the coupling constant the
representation of the form (one--gluon exchange approximation)
\be
\tilde
R(p;k|K)=ig^2[t^a_{(1)}][t^b_{(2)}]\gamma ^\mu _{(1)}\gamma ^\nu _{(2)}
D^{(0)ab}_{\mu \nu }(p-k),\label {7}
\ee
where  $t^a$  are  the  generators   of   the   gauge
transformations, $D_{\mu \nu }^{(0)ab}$ is a propagator of the
massless vector gluon, for which we use a standard expression
(in covariant gauge)[9]
\be
D_{\mu \nu }^{(0)ab}(q) = {\delta}^{ab}\frac{-1}{q^2+i\varepsilon }(g_{\mu \nu
}+(d^{(0)}-1)
\frac{q_{\mu }q_{\nu }}{q^2+i\varepsilon }). \label{8}
\ee
Here $d^{(0)}$ is a parameter, which fixes the gauge of gluon field.
Substituting the representation
(\ref{7}) for the function $\tilde R$ in R.H.S. of (\ref{6})
for the quark--antiquark interaction quasipotential we get [10,11]
\[
V_{q_1\bar{q}_2}(nM\vert \tilde{p}_1\sigma _{i_1}\tilde{p}_2\sigma _{i_2};
\tilde{k}_2\lambda _{j_2}\tilde{k}_1\lambda _{j_1}) = n_0\delta ^3(\vec{P}_M
-\vec{K}_M)
\delta _{\sigma _{f_1}}^{\lambda _{f_1}}\delta _{\sigma _{f_2}}^{\lambda _{f_2}}
\sum_{a}[t_{(1)}^{a}]
_{\sigma _{c_1}}^{\lambda _{c_1}}[t_{(2)}^{a}]_{\lambda _{c_2}}^{\sigma _{c_2}}\times\nonumber
\]
\[
\times\left[\bar{u}(\vec{p}_1\sigma _1)\gamma ^\mu u(\vec{k}_1\lambda _1)\bar{v}(\vec{k}_2\lambda _2)
\gamma _\mu v(\vec{p}_2\sigma _2)A^{(0)}(nM\vert \tilde{p}_1\tilde{p}_2;\tilde{k}_1
\tilde{k}_2)\right.+ \nonumber
\]
\be
+\left.(d^{(0)}-1)\bar{u}(\vec{p}_1\sigma _1)(n\gamma )u(\vec{k}_1\lambda _1)\bar{v}(\vec{k}_2\lambda _2)
(n\gamma )v(\vec{p}_2\sigma _2)B^{(0)}(nM\vert \tilde{p}_1\tilde{p}_2;\tilde{k}_1
\tilde{k}_2)\right]. \label{9}
\ee
Scalar functions $A^{(0)}$ and $B^{(0)}$ describing the properties
of quark-quark interactions in QCD in the given approximation are
defined with the help of the following integrals
\ber
\lefteqn{A^{(0)}(nM\mid \tilde{p}_1\tilde{p}_2;\tilde{k}_1
\tilde{k}_2) = }\nonumber\\
&=& \frac{g^2}{(2\pi )^5}\int_{-\infty }^{\infty }d\alpha \int_{-\infty
}^{\infty }d\beta
(\frac{1}{\Delta /2+\alpha -i\varepsilon }+\frac{1}{\Delta /2-\alpha
-i\varepsilon })\times \nonumber\\
&\times &(\frac{1}{\Delta '/2+\beta -i\varepsilon }+\frac{1}{\Delta
'/2-\beta -i\varepsilon })
\frac{1}{[p-k-(\alpha -\beta )n]^2+i\varepsilon },\ \label{10}
\eer
\ber
\lefteqn{B^{(0)}(nM\mid \tilde{p}_1\tilde{p}_2;\tilde{k}_1
\tilde{k}_2) = }\nonumber\\
&=& \frac{g^2}{(2\pi )^5}\int_{-\infty }^{\infty }d\alpha \int_{-\infty
}^{\infty }d\beta
(\frac{1}{\Delta /2+\alpha -i\varepsilon }+\frac{1}{\Delta /2-\alpha
-i\varepsilon })\times \nonumber\\
&\times &(\frac{1}{\Delta '/2+\beta -i\varepsilon }+\frac{1}{\Delta
'/2-\beta -i\varepsilon })
\frac{(\alpha -\beta )^2-(\Delta -\Delta ')^2/4}{([p-k-(\alpha -\beta
)n]^2+i\varepsilon )^2},\ \label{11}
\eer

The integrals defining the functions $A^{(0)}$ and $B^{(0)}$ may explicitly be
calculated. Here we shall present the calculation results for
the special evolution gauge (Markov--Yukawa gauge), where the
normal vector $n$ is directed along the total momentum of the
system, and for the case when the quark and antiquark masses are
equal: $m_1 = m_2 = m,\ \ p_{1\perp } = -p_{2\perp } = p_{\perp },\ \
k_{1\perp }=
-k_{2\perp } = k_{\perp }$, (in Appendix A one can
find the expressions for the functions $A^{(0)}$ and $B^{(0)}$ in an arbitrary
gauge and when quark and antiquark masses are not equal),
\ber
\lefteqn{A^{(0)}(M\mid p_{\perp };k_{\perp }) = }\label{12}\\
&=& \frac{g^2}{(2\pi )^3}\cdot\frac{1}{\sqrt{-(p_{\perp }-k_{\perp })^2}
(\sqrt{m^2-p_{\perp }^2}+\sqrt{m^2-k_{\perp }^2}+
\sqrt{-(p_{\perp }-k_{\perp })^2}-M)},
\nonumber
\eer
\ber
\lefteqn{B^{(0)}(M\mid p_{\perp };k_{\perp }) = }\label{13}\\
&=&
\frac{g^2}{2(2\pi )^3}[(\sqrt{m^2-p_{\perp }^2}+
\sqrt{m^2-k_{\perp }^2}-M)\times \nonumber\\
&\times &\frac{1}{\sqrt{-(p_{\perp }-k_{\perp })^2}(\sqrt{m^2-p_{\perp
}^2}+
\sqrt{m^2-k_{\perp }^2}+\sqrt{-(p_{\perp }-k_{\perp })^2}-M)^2}+
\nonumber\\
&+&\frac{(\sqrt{m^2-p_{\perp }^2}-\sqrt{m^2-k_{\perp }^2})^2}
{-(p_{\perp }-k_{\perp })^2(\sqrt{m^2-p_{\perp }^2}+\sqrt{m^2-k_{\perp
}^2}+
\sqrt{-(p_{\perp }-k_{\perp })^2}-M)^2}+ \nonumber\\
&+&\frac{(\sqrt{m^2-p_{\perp }^2}-\sqrt{m^2-k_{\perp }^2})^2}
{(\sqrt{-(p_{\perp }-k_{\perp })^2})^3(\sqrt{m^2-p_{\perp }^2}+
\sqrt{m^2-k_{\perp }^2}+\sqrt{-(p_{\perp }-k_{\perp })^2}-M)}],\nonumber
\eer
The calculated functions $A^{(0)}$ and $B^{(0)}$
correspond to the one--gluon exchange approximation. Note that the
function $A^{(0)}$ on the energy shell
\[
M = 2\sqrt{m^2-p_{\perp }^2}= 2\sqrt{m^2-k_{\perp }^2}
\]
takes the following form
\[
A^{(0)}(M\mid p_{\perp };k_{\perp })\mid _{on\ \ shell}\quad =
\frac{g^2}{(2\pi )^3}
\cdot\frac{1}{-(p_{\perp }-k_{\perp })^2}.
\]
The function $B^{(0)}$ turns into zero on the energy shell.

\section{Account of Infrared Singularities of Gluon Green's
Functions in QCD}

Nowadays there are many works, where one can find well
grounded arguments in favour of the singular infrared behaviour
$M^2/(k^2)^2$ for gluon Green's functions in QCD (see, for instance,
review [12] and references in it). In particular it is well
known, that the linear growth of the potential of
quark--antiquark interaction, which agrees with the experimental
data on quarkonium spectroscopy, corresponds to the static limit
of the diagram for one dressed gluon exchange, where the
propagator has the mentioned infrared asymptotics. Namely this
correspondence was in essence the very first and main argument
in favour of the assumption on such singular infrared
behaviour of the total one--particle gluon Green's
function. Further studies of the QCD structure allowed one to
make an important conclusion that infrared asymptotic $M^2/k^4$
of the gluon propagator yields a self--consistent
description of the QCD infrared region. This result and success
of the potential model in describing heavy quarkonium
spectroscopy with the quark interaction potential in the form of
a sum of a Coulomb--like term and the one linearly growing with
distance, make us think that we will obtain a sufficiently good
approximation for the total one--particle gluon Green's function
if we present it in the following form
\be
D_{\mu \nu } = D_{\mu \nu }^{(0)}(k) + D_{\mu \nu }^{(1)}(k),\label{15}
\ee
where $D_{\mu \nu }^{(0)}(k)$ determines the ultraviolet behaviour of the
gluon propagator and coincides with the free gluon Green's function,
and $D_{\mu \nu }^{(1)}(k)$ describes the singular infrared asymptotics
mentioned above so that
\ber
D_{\mu \nu }(k)& =& D_{\mu \nu }^{(0)}(k),\qquad k^2\rightarrow \infty
,\nonumber\\
D_{\mu \nu }(k)& =& D_{\mu \nu }^{(1)}(k),\qquad k^2\rightarrow 0.\nonumber
\eer

In the present Section we will show in the framework of
the single--time reduction method, what changes happen to the
two--quark interaction quasipotential in QCD if the infrared
singularities of the gluon propagator are taken into
consideration. A more accurate approximation we are going to make here,
consists in the fact that we intend to use in (\ref{7}) a total
one--particle gluon Green's function instead of a free one.
For the total Green's function we shall use the representation
(\ref{15}),
where $D_{\mu \nu }^{(0)}$ is defined above with Eq.(\ref{8}). And for
$D_{\mu \nu }^{(1)}$ we take the
representation from [12], which follows from the investigations of
infrared structure of QCD,
\be
D_{\mu \nu }^{(1)}(q) = \frac{\kappa ^2}{(q^2+i\varepsilon)^2 }(g_{\mu \nu
}+(d^{(1)}-1)
\frac{q_{\mu }q_{\nu }}{q^2+i\varepsilon }), \label{16}
\ee
where $d^{(1)}$ is a parameter, which in general does not
coincide with $d^{(0)}$. In the literature on infrared problem in QCD
we can find the ideas to consider the parameters $d^{(0)}$ and $d^{(1)}$
different in values. We shall come back to this important
problem a little bit later.

Substituting the expression for $D_{\mu \nu }$ from (\ref{15}) into the
R.H.S. of (\ref{7}) instead of $D_{\mu \nu }^{(0)}$ and taking into account
(\ref{8}) and (\ref{16}), for
the quark--antiquark interaction quasipotential we obtain
\[
V_{q_1\bar{q}_2}(nM\mid \tilde{p}_1\sigma _{i_1}\tilde{p}_2\sigma _{i_2};
\tilde{k}_2\lambda _{j_2}\tilde{k}_1\lambda _{j_1}) = n_0\delta
^3(\vec{P}_M
-\vec{K}_M)\delta _{\sigma _{f_1}}^{\lambda _{f_1}}\delta _{\sigma
_{f_2}}^{\lambda _{f_2}}\sum_{a}[t_{(1)}^{a}]
_{\sigma _{c_1}}^{\lambda _{c_1}}[t_{(2)}^{a}]_{\lambda _{c_2}}^{\sigma
_{c_2}}\nonumber
\]
\[
\times[\bar{u}(\vec{p}_1\sigma _1)\gamma ^\mu u(\vec{k}_1\lambda
_1)\bar{v}(\vec{k}_2\lambda _2)
\gamma _\mu v(\vec{p}_2\sigma _2)A(nM\mid
\tilde{p}_1\tilde{p}_2;\tilde{k}_1
\tilde{k}_2)+ \nonumber
\]
\be
+\bar{u}(\vec{p}_1\sigma _1)(n\gamma )u(\vec{k}_1\lambda
_1)\bar{v}(\vec{k}_2\lambda _2)
(n\gamma )v(\vec{p}_2\sigma _2)B(nM\mid \tilde{p}_1\tilde{p}_2;\tilde{k}_1
\tilde{k}_2)], \label{17}
\ee
where
\ber
\lefteqn{A(nM\mid \tilde{p}_1\tilde{p}_2;\tilde{k}_1\tilde{k}_2) = }
\nonumber\\
& & = A^{(0)}(nM\mid \tilde{p}_1\tilde{p}_2;\tilde{k}_1\tilde{k}_2)
+A^{(1)}(nM\mid \tilde{p}_1\tilde{p}_2;\tilde{k}_1\tilde{k}_2),
\label{18}
\eer
\ber
B(nM\mid \tilde{p}_1\tilde{p}_2;\tilde{k}_1\tilde{k}_2)&=&(d^{(0)}-1)
B^{(0)}(nM\mid \tilde{p}_1\tilde{p}_2;\tilde{k}_1\tilde{k}_2)+
\nonumber\\ &+& (d^{(1)}-1)
B^{(1)}(nM\mid \tilde{p}_1\tilde{p}_2;\tilde{k}_1\tilde{k}_2),
\label{19}
\eer
Scalar functions $A^{(1)}$ and $B^{(1)}$  are defined with the
help of the following integrals [8]
\ber
\lefteqn{A^{(1)}(nM\mid \tilde{p}_1\tilde{p}_2;\tilde{k}_1
\tilde{k}_2) = }\nonumber\\
&=& \frac{-(g\kappa )^2}{(2\pi )^5}\int_{-\infty }^{\infty }d\alpha
\int_{-\infty }^{\infty }d\beta
(\frac{1}{\Delta /2+\alpha -i\varepsilon }+\frac{1}{\Delta /2-\alpha
-i\varepsilon })\times \nonumber\\
&\times &(\frac{1}{\Delta '/2+\beta -i\varepsilon }+\frac{1}{\Delta
'/2-\beta -i\varepsilon })
\frac{1}{([p-k-(\alpha -\beta )n]^2+i\varepsilon )^2},\ \label{20}
\eer
\ber
\lefteqn{B^{(1)}(nM\mid \tilde{p}_1\tilde{p}_2;\tilde{k}_1
\tilde{k}_2) = }\nonumber\\
&=& \frac{-(g\kappa )^2}{(2\pi )^5}\int_{-\infty }^{\infty }d\alpha
\int_{-\infty }^{\infty }d\beta
(\frac{1}{\Delta /2+\alpha -i\varepsilon }+\frac{1}{\Delta /2-\alpha
-i\varepsilon })\times \nonumber\\
&\times &(\frac{1}{\Delta '/2+\beta -i\varepsilon }+\frac{1}{\Delta
'/2-\beta -i\varepsilon })
\frac{(\alpha -\beta )^2-(\Delta -\Delta ')^2/4}{([p-k-(\alpha -\beta
)n]^2+i\varepsilon )^3}.\ \label{21}
\eer

The integrals defining the functions $A^{(1)}$ and $B^{(1)}$ may explicitly be
calculated. Here we shall present as in previous Section the
calculation results for the special Markov--Yukawa evolution gauge
and for the case when the quark and antiquark masses are
equal: $m_1 = m_2 = m,\ \ p_{1\perp } = -p_{2\perp } = p_{\perp },\ \
k_{1\perp }=
-k_{2\perp } = k_{\perp }$, (in Appendix A one can
find the expressions for the functions $A^{(1)}$ and $B^{(1)}$ in general
case),
\ber
\lefteqn{A^{(1)}(M\mid p_{\perp };k_{\perp }) = }\label{22}\\
&=& \frac{(g\kappa )^2}{(2\pi )^3}\cdot\frac{1}{-2(p_{\perp }-k_{\perp
})^2}\times
\nonumber\\
&\times &\left[\frac{1}{(\sqrt{m^2-p_{\perp }^2}+
\sqrt{m^2-k_{\perp }^2}+\sqrt{-(p_{\perp }-k_{\perp
})^2}-M)^2}\right.+\nonumber\\
&+&\left.\frac{1}{\sqrt{-(p_{\perp }-k_{\perp })^2}
(\sqrt{m^2-p_{\perp }^2}+\sqrt{m^2-k_{\perp }^2}+
\sqrt{-(p_{\perp }-k_{\perp })^2}-M)}\right],\nonumber
\eer
\newpage
\ber
\lefteqn{B^{(1)}(M\mid p_{\perp };k_{\perp }) = }\label{23}\\
&=& -\frac{(g\kappa )^2}{(2\pi )^3}\cdot
\frac{1}{4(\sqrt{-(p_{\perp }-k_{\perp })^2})^3}\times \nonumber\\
&\times &\left[\frac{(\sqrt{m^2-p_{\perp }^2}+\sqrt{m^2-k_{\perp }^2}-M)^2}
{(\sqrt{m^2-p_{\perp }^2}+\sqrt{m^2-k_{\perp }^2}+\sqrt{-(p_{\perp
}-k_{\perp })^2}
-M)^3}\right. - \nonumber\\
&-&\frac{3}{2}\cdot\frac{\sqrt{m^2-p_{\perp }^2}+\sqrt{m^2-k_{\perp }^2}-M}
{(\sqrt{m^2-p_{\perp }^2}+\sqrt{m^2-k_{\perp }^2}+\sqrt{-(p_{\perp
}-k_{\perp })^2}
-M)^2} - \nonumber\\
&-&\frac{(\sqrt{m^2-p_{\perp }^2}-\sqrt{m^2-k_{\perp }^2})^2}
{(\sqrt{m^2-p_{\perp }^2}+\sqrt{m^2-k_{\perp }^2}+\sqrt{-(p_{\perp
}-k_{\perp })^2}
-M)^3} - \nonumber\\
&-& \frac{3(\sqrt{m^2-p_{\perp }^2}-\sqrt{m^2-k_{\perp }^2})^2}
{2\sqrt{-(p_{\perp }-k_{\perp })^2}(\sqrt{m^2-p_{\perp }^2}+
\sqrt{m^2-k_{\perp }^2}+\sqrt{-(p_{\perp }-k_{\perp })^2}-M)^2}-
\nonumber\\
&-&\left. \frac{3(\sqrt{m^2-p_{\perp }^2}-\sqrt{m^2-k_{\perp }^2})^2}
{2(\sqrt{-(p_{\perp }-k_{\perp })^2})^2(\sqrt{m^2-p_{\perp }^2}+
\sqrt{m^2-k_{\perp }^2}+\sqrt{-(p_{\perp }-k_{\perp })^2}-M)}\right].
\nonumber
\eer
The functions $A^{(1)}$ and $B^{(1)}$ originated from the infrared singular
part of the total gluon propagator. Note that the function
$A^{(1)}$ on the energy shell
$M = 2\sqrt{m^2-p_{\perp }^2}= 2\sqrt{m^2-k_{\perp }^2}$ takes the
following form
\[
A^{(1)}(M\mid p_{\perp };k_{\perp })\mid _{on\ \ shell}\quad =
\frac{(g\kappa )^2}{(2\pi )^3}
\cdot\frac{1}{(p_{\perp }-k_{\perp })^4}.
\]
Similar to the function $B^{(0)}$ the function $B^{(1)}$ turns into zero
on the energy shell.

\section{Configuration Space and Local Approxi\-ma\-tions}

Dynamic functions $A$ and $B$, defined with formulae
(\ref{18},\ref{19}),
characterize the interaction of two quarks in QCD in
one--gluon exchange approximation with an account of the infrared
singularities of the total gluon propagator. In order to analyze
the dynamic functions in the configuration space it will
be more convenient to go over to new variables in these
functions which are defined in the following way
\[
{\tilde{p}}_i = L(n) {\ovc{\tilde{p}}}_i,\qquad
{\tilde{k}}_i = L(n) {\ovc{\tilde{k}}}_i,
\]
where $L(n)$ is the matrix of a pure Lorentz transformation with
elements
\[
L(n)_0^\mu = L(n)_\mu ^0 = n^\mu ,\qquad L(n)_j^i = \delta _j^i-
(1+n_0)^{-1}n^in_j.
\]
As is easily seen, in this case the variables defined above are
transformed to the form
\be
{\ovc{p}}_{1\perp } = -{\ovc{p}}_{2\perp } = {\ovc{p}}_{\perp } =
(0,\ovc{\vec{p}}),
\qquad {\ovc{k}}_{1\perp } = -{\ovc{k}}_{2\perp } = {\ovc{k}}_{\perp } =
(0,\ovc{\vec{k}}),
\label{25}
\ee
\[
\ovc{\vec{p}} = {\ovc{\vec{p}}}_1 = -{\ovc{\vec{p}}}_2,\qquad
\ovc{\vec{k}} = {\ovc{\vec{k}}}_1 = -{\ovc{\vec{k}}}_2,
\]
in this,
\be
p_{\perp }^2 = ({\ovc{p}}_{\perp })^2 = -(\ovc{\vec{p}})^2,\qquad
k_{\perp }^2 = ({\ovc{k}}_{\perp })^2 = -(\ovc{\vec{k}})^2.\label{26}
\ee
In the terms of new variables and account of transformation
properties of bispinors we obtain for the spinor structure of
the interaction potential the following expression
\ber
u^+(\ovc{\vec{p}}\sigma _1)u(\ovc{\vec{k}}\lambda
_1)v^+(-\ovc{\vec{k}}\lambda _2)
v(-\ovc{\vec{p}}\sigma _2)[A(M\mid \ovc{\vec{p}};\ovc{\vec{k}})+
B(M\mid \ovc{\vec{p}};\ovc{\vec{k}})]- \nonumber\\
-\bar{u}(\ovc{\vec{p}}\sigma _1)\vec{\gamma }u(\ovc{\vec{k}}\lambda _1)
\bar{v}(-\ovc{\vec{k}}\lambda _2)\vec{\gamma }v(-\ovc{\vec{p}}\sigma _2)
A(M\mid \ovc{\vec{p}};\ovc{\vec{k}}),\qquad \label{27}
\eer
where the expressions for the functions $A$ and $B$ in the terms of
new variables are simply derived from the relevant formulae
(\ref{12},\ref{13}), (\ref{22},\ref{23}) through trivial substitutions
(\ref{25},\ref{26}).
From explicit expressions (\ref{12},\ref{13},\ref{22},,\ref{23})
it is seen that the
dynamic functions $A$ and $B$, defining the properties of the
interaction potential for quark and antiquark, are non--local
functions depending on the total energy of the quark--antiquark
system. This result is a consequence of a
consistent relativistic consideration of the two body problem
in the framework of the local quantum field theory. For the
quark--antiquark configuration, where the conditions
$\ovc{\vec{p}\,^2}/m^2<<1,\quad \ovc{\vec{k}^2}/m^2<<1$ are
fulfilled, one can approximate the dynamic functions $A$ and
$B$ by the local functions with good accuracy.
For instance for the functions $A^{(1)}$ and $B^{(1)}$ we find that
\ber
\lefteqn{A^{(1)}(M\mid p_{\perp };k_{\perp }) \cong A^{(1)}(\varepsilon
;\mid \ovc{\vec{p}}-
\ovc{\vec{k}}\mid ) = } \label{28}\\
&=& \frac{(g\kappa )^2}{2(2\pi )^3\mid \ovc{\vec{p}}-\ovc{\vec{k}}\mid ^2}
\left[\frac{1}{(\mid \ovc{\vec{p}}-\ovc{\vec{k}}\mid -\varepsilon )^2} +
\frac{1}
{{\mid \ovc{\vec{p}}-\ovc{\vec{k}}\mid }(\mid \ovc{\vec{p}}-
\ovc{\vec{k}}\mid -\varepsilon )}\right], \nonumber
\eer
\ber
\lefteqn{B^{(1)}(M\mid p_{\perp };k_{\perp }) \cong B^{(1)}(\varepsilon
;\mid \ovc{\vec{p}}-
\ovc{\vec{k}}\mid ) = } \label{29}\\
&=& \frac{-(g\kappa )^2}{4(2\pi )^3\mid \ovc{\vec{p}}-\ovc{\vec{k}}\mid ^3}
\left[\frac{\varepsilon ^2}{(\mid \ovc{\vec{p}}-\ovc{\vec{k}}\mid
-\varepsilon )^3} + \frac{3}{2}\cdot
\frac{\varepsilon }{(\mid \ovc{\vec{p}}-\ovc{\vec{k}}\mid -\varepsilon
)^2}\right], \nonumber
\eer
where $\varepsilon = M-2m$ is the binding energy of the quark--antiquark
system. We obtain also corresponding  local
approximations for the functions $A^{(0)}$ and $B^{(0)}$ [10]
\be
A^{(0)}(\varepsilon ;\mid \ovc{\vec{p}}-\ovc{\vec{k}}\mid =\frac{g^2}{(2\pi
)^3}
\cdot\frac{1}{\mid \ovc{\vec{p}}-\ovc{\vec{k}}\mid (\mid \ovc{\vec{p}}-
\ovc{\vec{k}}\mid -\varepsilon )}, \label{30}
\ee
\be
B^{(0)}(\varepsilon ;\mid \ovc{\vec{p}}-\ovc{\vec{k}}\mid =\frac{g^2}{(2\pi
)^3}
\cdot\frac{-\varepsilon }{2\mid \ovc{\vec{p}}-\ovc{\vec{k}}\mid (\mid
\ovc{\vec{p}}-
\ovc{\vec{k}}\mid -\varepsilon )^2}. \label{31}
\ee

In the configuration space, to which we pass through the
Fourier transformation
\be
A(\varepsilon ;r)=\int d\ovc{\vec{q}}exp(i\ovc{\vec{q}}\ovc{\vec{x}})
A(\varepsilon ;\mid \ovc{\vec{q}}\mid ),\quad r\equiv \mid
\ovc{\vec{x}}\mid \label{32}
\ee
(and a similar integral for the function $B$), local
energy--dependent potentials will correspond to the functions
(\ref{28}--\ref{31}). The expressions for the functions $A^{(0)}$ and
$B^{(0)}$ have the form [10]
\[
A^{(0)}(\varepsilon ;r)=\frac{g^2}{4\pi r}\cdot\frac{2}{\pi }\cdot
a(\bar\varepsilon r),
\quad B^{(0)}(\varepsilon ;r)=\frac{g^2}{4\pi r}\cdot\frac{\bar\varepsilon
r}{\pi }\cdot
b(\bar\varepsilon r)
\]
in the case of a negative binding energy $\varepsilon = -\bar\varepsilon
<0, \bar\varepsilon >0$ and
\ber
A^{(0)}(\varepsilon ;r)=\frac{g^2}{4\pi r}\cdot\frac{2}{\pi }[\pi
e^{i\varepsilon r}- a(\varepsilon r)],
\nonumber\\
B^{(0)}(\varepsilon ;r)=-\frac{g^2}{4\pi r}\cdot\frac{\varepsilon r}{\pi
}\cdot[b(\varepsilon r)+
i\pi e^{i\varepsilon r}] \nonumber
\eer
in the case of positive binding energy $\varepsilon >0$, where
\[
a(x) = ci(x)sin(x) - si(x)cos(x),\quad b(x) = -ci(x)cos(x)-si(x)sin(x),
\]
$ci(x)$ and $si(x)$ are integral cosines and sines.

As can easily be seen, integral (\ref{32}), determining the functions
$A^{(1)}$ and $B^{(1)}$ in the configuration space, diverge. With the help of
the standard regularization procedures the divergent part in the
integral for the functions $A^{(1)}$ and $B^{(1)}$ may easily be singled out.
As a result we obtain
\ber
A^{(1)}(\varepsilon ;r)=-\frac{\kappa ^2}{2\bar\varepsilon
^2}\cdot\frac{g^2}{4\pi r}\cdot
\frac{2}{\pi }\cdot\bar\varepsilon r[b(\bar\varepsilon r)-b(\mu r)],\quad
\mu \rightarrow 0 \label{36}\\
B^{(1)}(\varepsilon ;r)= \frac{\kappa ^2}{4\bar\varepsilon
^2}\cdot\frac{g^2}{4\pi r}\cdot
\frac{\bar\varepsilon r}{\pi }[\bar\varepsilon ra(\bar\varepsilon
r)-1-b(\bar\varepsilon r)+b(\mu r)],\quad \mu \rightarrow 0
\label{37}
\eer
in the case of a negative binding energy and
\ber
A^{(1)}(\varepsilon ;r)&=&\frac{\kappa ^2}{2\varepsilon
^2}\cdot\frac{g^2}{4\pi r}\cdot
\frac{2}{\pi }\cdot \varepsilon r[b(\varepsilon r)+i\pi e^{i\varepsilon
r}-b(\mu r)],\quad \mu \rightarrow 0 \label{38}\\
B^{(1)}(\varepsilon ;r)&=& \frac{\kappa ^2}{4\varepsilon
^2}\cdot\frac{g^2}{4\pi r}\cdot
\frac{\varepsilon r}{\pi }[(i+\varepsilon r)\pi e^{i\varepsilon
r}-\varepsilon ra(\varepsilon r)+ \nonumber\\
&+& 1+b(\varepsilon r)-b(\mu r)],\quad \mu \rightarrow 0 \label{39}
\eer
in the case of a positive binding energy. The function $b(x)$
has a logarithmic singularity at zero and, as can easily be
seen, one and the same infinite constant $b(0)$ is present in
expressions (\ref{36}---\ref{39}). Here we find one very important
circumstance, which consists in the following. From the spinor
structure of (\ref{27}) it follows that the spin--independent part of
the interaction quasipotential is determined with a linear combination
of the dynamic functions $A+B$, which will be presented in the
form
\[
A+B=V^{(0)}+V^{(1)}\equiv V,
\]
where
\ber
V^{(0)}=A^{(0)}+(d^{(0)}-1)B^{(0)}, \label{40}\\
V^{(1)}=A^{(1)}+(d^{(1)}-1)B^{(1)}. \label{41}
\eer
It turns out that there exists a special gauge $d^{(1)}=-3$,
where the infinities mentioned above are
canceled, and we come to the finite result for the function
$V^{(1)}$
\be
V_A^{(1)}(\varepsilon ;r)\equiv V^{(1)}(\varepsilon ;r)\mid
_{d^{(1)}=-3}\quad=
\frac{\kappa ^2}{\bar\varepsilon ^2}\cdot\frac{g^2}{4\pi r}\cdot
\frac{\bar\varepsilon r}{\pi }[1-\bar\varepsilon ra(\bar\varepsilon r)],
\label{42}
\ee
in the case of the negative binding energy and in the case of
this energy being positive we have
\be
V_A^{(1)}(\varepsilon ;r)=-\frac{\kappa ^2}{\varepsilon
^2}\cdot\frac{g^2}{4\pi r}\cdot
\frac{\varepsilon r}{\pi }[1-\varepsilon ra(\varepsilon r)+\pi \varepsilon
re^{i\varepsilon r}]. \label{43}
\ee
This remarkable result of the cancellation of divergences,
which leads to the finite function $V^{(1)}$, seems to be connected
with the property of gauge $d^{(1)}=-3$ discussed in [12], which
manifests itself in the fact, that in this gauge gluon Green's
function is transverse in the coordinate space, which, in its
turn, guarantees the existence of the static color charge field.
We shall present also the expression for the function
$V_A^{(1)}$ in the momentum space
\ber
\lefteqn{V_A^{(1)}(M\mid p_{\perp };k_{\perp }) = }\label{44}\\
&=& \frac{(g\kappa )^2}{(2\pi )^3}\cdot
\frac{1}{\sqrt{-(p_{\perp }-k_{\perp })^2}}\times \nonumber\\
&\times &\left[\frac{1}
{(\sqrt{m^2-p_{\perp }^2}+\sqrt{m^2-k_{\perp }^2}+\sqrt{-(p_{\perp
}-k_{\perp })^2}
-M)^3}\right. - \nonumber\\
&-&\frac{(\sqrt{m^2-p_{\perp }^2}-\sqrt{m^2-k_{\perp }^2})^2}
{(\sqrt{-(p_{\perp }-k_{\perp })^2})^2(\sqrt{m^2-p_{\perp }^2}+
\sqrt{m^2-k_{\perp }^2}+\sqrt{-(p_{\perp }-k_{\perp })^2}-M)^3} -
\nonumber\\
&-& \frac{3(\sqrt{m^2-p_{\perp }^2}-\sqrt{m^2-k_{\perp }^2})^2}
{2(\sqrt{-(p_{\perp }-k_{\perp })^2})^3(\sqrt{m^2-p_{\perp }^2}+
\sqrt{m^2-k_{\perp }^2}+\sqrt{-(p_{\perp }-k_{\perp })^2}-M)^2}-
\nonumber\\
&-&\left. \frac{3(\sqrt{m^2-p_{\perp }^2}-\sqrt{m^2-k_{\perp }^2})^2}
{2(\sqrt{-(p_{\perp }-k_{\perp })^2})^4(\sqrt{m^2-p_{\perp }^2}+
\sqrt{m^2-k_{\perp }^2}+\sqrt{-(p_{\perp }-k_{\perp })^2}-M)}\right].
\nonumber
\eer
The corresponding local approximation for the function
$V_A^{(1)}$, analogous to formulae (\ref{28}--\ref{31}), has the form
\be
V_A^{(1)}(\varepsilon ;\mid \ovc{\vec{p}}-\ovc{\vec{k}}\mid =\frac{(g\kappa
)^2}{(2\pi )^3}
\cdot\frac{1}{\mid \ovc{\vec{p}}-\ovc{\vec{k}}\mid (\mid \ovc{\vec{p}}-
\ovc{\vec{k}}\mid -\varepsilon )^3}. \label{45}
\ee
One may also get convinced that the expressions in the R.H.S.of
formulae (\ref{42}) and (\ref{43}) can be obtained through the Fourier
transformation of function (\ref{45}).

It will be interesting to study asymptotic properties of the
function $V_A^{(1)}$ in the region of large and
small distances. Using the known asymptotic expansions for
$ci(x)$ and $si(x)$ [13], we obtain from (\ref{42}) and (\ref{43});

a) the binding energy is negative $\varepsilon =-\bar\varepsilon <0,\quad
\bar\varepsilon >0$:
\be
V_A^{(1)}(\varepsilon ;r)=\frac{\kappa ^2}{\bar\varepsilon ^2}\cdot
\frac{2\alpha \bar\varepsilon }
{\pi (\bar\varepsilon r)^2}\left[1-\frac{12}{(\bar\varepsilon
r)^2}+O(\frac{1}
{(\bar\varepsilon r)^4})\right],\quad r>>\frac{1}{\bar\varepsilon
},\label{46}
\ee
\be
V_A^{(1)}(\varepsilon ;r)=\frac{\kappa ^2}{\bar\varepsilon ^2}\cdot
\frac{\alpha \bar\varepsilon }
{\pi }[1-\frac{\pi }{2}\bar\varepsilon r-(\bar\varepsilon r)^2(ln(\gamma
\bar\varepsilon r)-1)+O((\bar\varepsilon r)^3)],
\quad r<<\frac{1}{\bar\varepsilon }, \label{47}
\ee

b) the binding energy is positive $\varepsilon >0$:
\be
V_A^{(1)}(\varepsilon ;r)=-\alpha \kappa ^2re^{i\varepsilon
r}\left[1+O(\frac{1}{(\varepsilon r)^3})\right],\quad
r>>\frac{1}{\varepsilon }, \label{48}
\ee
\be
V_A^{(1)}(\varepsilon ;r)=-\frac{\kappa ^2}{\varepsilon ^2}\cdot
\frac{\alpha \varepsilon }
{\pi }[1+\frac{\pi }{2}\varepsilon r-(\varepsilon r)^2(ln(\gamma
\varepsilon r)-1-i\pi )+O((\varepsilon r)^3)],
\label{49}\\
r<<\frac{1}{\varepsilon }
\ee
where we put $\alpha =g^2/4\pi $.
Hence, in the discrete spectrum (the binding energy is negative)
in the range of large distances the function $V_A^{(1)}$
decreases at the infinity more rapidly than the Coulomb one,
which coincides with the corresponding asymptotic behaviour of
the function $V^{(0)}(\varepsilon ;r)$. In the region of small
distances in
the discrete spectrum the behaviour of the function $V_A^{(1)}$ differs
greatly from the behaviour of $V^{(0)}$ which has a Coulomb
singularity at zero. The function $V_A^{(1)}(\varepsilon ,r)$ is inversely
proportional to the binding energy with the proportionality
coefficient equal to $\alpha \kappa ^2/\pi $, at zero,i.e.,
\[
V_A^{(1)}(\varepsilon ;r)\mid _{r=0}\quad=\frac{\alpha \kappa ^2}{\pi
\bar\varepsilon },\quad \varepsilon =-\bar\varepsilon <0.
\]
The same difference in the behaviour of the functions $V_A^{(1)}$ and
$V^{(0)}$ in the region of small distances holds in the case of a
continuous spectrum, when the binding energy is positive,
in this case
\[
V_A^{(1)}(\varepsilon ;r)\mid _{r=0}\quad=-\frac{\alpha \kappa ^2}{\pi
\varepsilon },\quad \varepsilon >0.
\]

Here we shall make a remark connected with the following fact.
In ref.[12] it has been shown that the singular structure of
the gluon propagator within the framework of the dimensional
regularization used in the given paper, depends on the way of
limiting transition to the physical dimensionality $n=4$ of the
space-time. In particular, it has been noted, that a self consistent
description of the ghost and gluon Green's functions
fixes such a transition to the physical dimensionality of the
space where the singularity structure of the gluon propagator
in the infrared region has the form
\be
D_{\mu \nu }(k)=D_{\mu \nu }^{(1)}(k)+D_{\mu \nu }^{(2)}(k),\quad
k^2\rightarrow 0,\label{50}
\ee
where
\be
D_{\mu \nu }^{(2)}(k)=-2\pi ^2\kappa ^2i\delta ^{(4)}(k)g_{\mu \nu
}.\label{51}
\ee
One can easily guess that taking account of an additional term of
form (\ref{51}) with the help of the scheme presented above,
leads to the appearance of an additional term in the function $A$,
which will now be equal to
\[
A=A^{0)}+A^{(1)}+A^{(2)}.
\]
We obtain an explicit expression for $A^{(2)}$
\be
A^{(2)}(M\mid p_{\perp };k_{\perp })=-\frac{(g\kappa )^2}{(2\pi )^3}\cdot
\frac{2\pi n_0\delta ^{(3)}(\vec{p}_{\perp }-\vec{k}_{\perp })}
{2\sqrt{m^2-p_{\perp }^2}-M}, \label{52}
\ee
which takes a very simple form in the local limit
\be
A^{(2)}(\varepsilon ;\ovc{\vec{p}}-\ovc{\vec{k}})=\frac{(g\kappa )^2}{(2\pi
)^3}\cdot
\frac{2\pi }{\varepsilon }\delta
^{(3)}(\ovc{\vec{p}}-\ovc{\vec{k}}).\label{53}
\ee
In the configuration space a constant distance--independent term
\be
A^{(2)}(\varepsilon ;r)=\frac{\alpha \kappa ^2}{\pi \varepsilon }\label{54}
\ee
will correspond to function (\ref{53}). The function $V$ takes an even
simpler form if the additional term (\ref{54}) is taken into
consideration
\ber
\lefteqn{V_A(\varepsilon ;r)\equiv V_A^{(1)}(\varepsilon
;r)+A^{(2)}(\varepsilon ;r)=}\nonumber\\
&=&-\frac{\alpha \kappa ^2}{\pi }\cdot ra(\bar\varepsilon r),\quad
\varepsilon =-\bar\varepsilon <0,
\nonumber\\
&=&\frac{\alpha \kappa ^2}{\pi }\cdot r(a(\varepsilon r)-\pi
e^{i\varepsilon r}),\quad \varepsilon >0.\label{55}
\eer

The essential difference of the changed function $V_A$ from
function $V_A^{(1)}$ manifests itself in the fact that in the
expression
(\ref{55}) a correct transition to the limit of the zero binding
energy is allowed, and as can easily be seen, only in the limit
of the zero binding energy we come to the potential linearly
growing with distance. Besides from formula (\ref{55}) we get
\[
V_A(\varepsilon ;0)=0.
\]
It is obvious that the indicated differences may essentially
influence on the results of the data analysis for spectroscopy and
decays of quark systems. Because of the importance of the
circumstance we should also note here the changes which
occur in the asymptotic behaviour of the function $V_A$;

a) the binding energy is negative:
\be
V_A(\varepsilon ;r)=-\frac{\alpha \kappa ^2}{\pi \bar\varepsilon }
\left[1-\frac{2}{(\bar\varepsilon r)^2}+O(\frac{1}{(\bar\varepsilon
r)^4})\right],
\quad r>>\frac{1}{\bar\varepsilon },\label{56}
\ee
\be
V_A(\varepsilon ;r)=-\frac{\alpha \kappa ^2}{\pi }\cdot r
[\frac{\pi }{2}+(\bar\varepsilon r)(ln(\gamma \bar\varepsilon
r)-1)+O((\bar\varepsilon r)^2)],
\quad r<<\frac{1}{\bar\varepsilon }, \label{57}
\ee

b) the binding energy is positive:
\be
V_A(\varepsilon ;r)=-\alpha \kappa ^2re^{i\varepsilon
r}\left[1+O(\frac{1}{(\varepsilon r)})\right],\quad
r>>\frac{1}{\varepsilon }, \label{58}
\ee
\be
V_A(\varepsilon ;r)=-\frac{\alpha \kappa ^2}{\pi }\cdot r
[\frac{\pi }{2}-(\varepsilon r)(ln(\gamma \varepsilon r)-1-i\pi
)+O((\varepsilon r)^2)],r<<\frac{1}{\varepsilon }.
\label{59}
\ee

In the continuous spectrum in the range of large distances
the function $V_A^{(1)}$ has oscillations with the amplitude, linearly
growing with the distance with the proportionality coefficient
equal to $\alpha \kappa ^2$. In the range of large distances
in the continuous spectrum the function $V^{(0)}$ is characterized by
the same oscillations, as the function $V_A^{(1)}$, but with the
amplitude having two components: decreasing with the distance
according to the Coulomb law and constant which proportional to
the binding energy. More precisely we have
\[
V^{(0)}(\varepsilon ;r)=2\alpha e^{i\varepsilon
r}\left[\frac{1}{r}+\frac{d^{(0)}-1}{2i}
\cdot \varepsilon \right],r>>\frac{1}{\varepsilon }.
\]

One can see another interesting property, namely a weak
dependence of the interaction quasipotential on the choice of the
gauge at negative binding energy.

Contrary to the discrete spectrum in the case of the
continuous spectrum (at the positive binding energy), there is
quite a noticeable dependence of
the interaction quasipotential $V^{(0)}$ (without taking account of the infrared
singularities) on the gauge, which is characterized by the
presence of the "knot" points, where the potential $V^{(0)}$ has one
and the same value at any values of the gauge parameter
$d^{(0)}$. Such a peculiarity in the behaviour of the
interaction quasipotential is
conserved even one takes account of infrared singularities, in this
in the range of large distances the contribution of
infrared singularities to the interaction quasipotential for
quark and antiquark is decisive.

\section{Asymptotic Freedom and Quark--Quark\newline Forces in QCD}

In the previous Sections the single--time  reduction
technique of the Bethe--Salpeter formalism for two--fermion systems
[8] was applied to the problem of calculating the two--quark
 interaction   quasipotential   in   the    one--gluon    exchange
approximation  in  QCD.  In  this  approximation,  the   analytic
expressions  for   the   quasipotential   of
two--quark interactions were obtained,  allowing  explicitly
for the structure of the initial gauge model. It was shown that a
consistent relativistic consideration of  the  quark  interaction
problem allows to establish a nontrivial energy dependence of the
quark interaction potential. This  energy  dependence  gives  the
interaction potential quite unusual properties concerning  its
behavior in the configuration space. In particular, as  a  result
of the  energy  dependence  of  the  interaction  potential,  the
character of the forces changes qualitatively during the  transition
from the discrete spectrum (the region of the negative values  of
the binding energy) to  the  continuous  spectrum  (that  of  the
positive values  of  the  binding  energy).  Namely,  the  smooth
behavior of the interaction potential in the discrete spectrum
goes into the oscillations in  the  continuous  spectrum.

Using  the  ansatz  about  the  singular  behavior  of  a   gluon
propagator in the infrared region, we've  explored  how  infrared
singularities of gluon Green's functions affect the  behavior  of
quark--quark forces in quantum chromodynamics.

The  singular  behavior  of  gluon   Green's   functions   is   a
characteristic property of the  non--Abelian  gauge  model  under
consideration and originates from the nonperturbative research of the
infrared region  in  QCD  [12].  Another  peculiarity  of quantum
chromodynamics is the discovered asymptotic freedom of the model,
which is testified by the decrease of a running coupling constant
with the growth of a transferred momentum. This property
is established
by the perturbative analysis of QCD [14,15]. The asymptotic  freedom
allows to  calculate  things  in  perturbation  theory  at  small
distances, and to compare the results with experimentally measurable
quantities at  large  momentum  transfers  or  large  transversal
momenta.

In the present Section we will show in a  consistent
relativistic way how  the  asymptotic
freedom displays the character of quark--quark forces.

\subsection{Generalized Richardson's Parameterization and\newline
 Single--Time Formalism in Quantum Chromo\-dy\-na\-mics}

In calculations of the  quasipotential of a two--quark  interaction
which have been presented in previous Sections for  the
function  $\tilde  R$  in  the  R.H.S.  of  Eq. (\ref{6})   the
representation of the form (one--gluon exchange approximation)
\be
\tilde
R(p;k|K)=ig^2[t^a_{(1)}][t^b_{(2)}]\gamma ^\mu _{(1)}\gamma ^\nu _{(2)}
D^{ab}_{\mu \nu }(p-k),\label {61}
\ee
was  used,  where  $t^a$  are  the  generators   of   the   gauge
transformations, $D$ is the  gluon propagator, for which in  turn
the following ansatz was used:
\be
D^{ab}_{\mu \nu }(q)= \delta ^{ab}(D^{(0)}_{\mu \nu }(q)+
D^{(1)}_{\mu \nu }(q)).\label {62}
\ee
Here $D^{(1)}_{\mu \nu }$ determines the infrared behavior of the gluon
propagator, and $D^{(0)}_{\mu \nu }(q)$ coincides with the  free  gluon
Green's function. As usual, in formula (\ref{61})  summation  over
the repeating indices is assumed.  As it  turned  out,  already  the
level of the one--gluon exchange approximation  reveals  many
interesting features of the behavior of quark--quark  forces.
Some  of  these  features  were mentioned  above.  A
disadvantage of representation (\ref{61})  is  that  it  does  not
take into account the property of  asymptotic  freedom  discovered  in
quantum chromodynamics. This property, however, may easily be taken
into account if one uses for the function $\tilde R$ the following
representation:
\be
\tilde R(p;k|K)=i\alpha _s(q^2)[t^a_{(1)}][t^b_{(2)}]\gamma ^\mu _{(1)}\gamma ^\nu _{(2)}
D^{(0)ab}_{\mu \nu }(q),\quad q\equiv p-k\label{63},
\ee
where $\alpha _s(q^2)$ is an invariant charge, for  which  we  take
the expression that was obtained in QCD in the one--loop
approximation [14,15]
\be
\alpha _s(Q^2)=\frac{\alpha _s(\mu ^2)}{1+b\alpha _s(\mu ^2)ln(Q^2/\mu ^2)}+0(\alpha _s),\quad
Q^2\equiv -q^2,\label {64}
\ee
with $\alpha _s(\mu ^2)\equiv g^2/4\pi \equiv \alpha _s$ a physical coupling constant, and  the
$b$ parameter depending on the structure of the gauge group.  For
the group $SU_c(3)$ the $b$ parameter is equal to
\[
b=\frac{1}{12\pi }(33-2n_f),\nonumber
\]
with $n_f$ the  number  of  the  quark  flavors.  Instead  of  the
dimensional parameter $\mu ^2$, it would be  convenient  to  bring  in
another dimensional parameter, $\Lambda ^2$, through the relation
\[
ln \Lambda ^2=ln \mu ^2-\frac {1}{b\alpha _s(\mu ^2)}.\nonumber
\]
Then we have
\be
\alpha _s(Q^2)=\frac{1}{b
ln(Q^2/\Lambda ^2)}=\frac{4\pi }{(11-\frac{2}{3}n_f)
ln(Q^2/\Lambda ^2)}.\label {65}
\ee
The applicability region of the  one--loop approximation for
an invariant charge is established from its derivation.
This is  the  region  of  large  $Q^2:Q^2\gg \Lambda ^2$.  Therefore,
strictly speaking, representation (\ref{63}) for the  function
$\tilde R$ with the expression for the invariant  charge  in
form  (\ref{65})  should  be  considered  as   an  asymptotic
representation which works in the region of large $Q^2$. Here
the situation differs from (\ref{61}): representation (\ref{63}),
although describing the ultraviolet behavior of the function
$\tilde R$, is quite unfit for the description of the things in the
infrared region. However, a simple  trick  invented by Richardson
[16] is available to sew these two asymptotics.  The Richardson's
parameterization looks like
\be
\tilde
R(p;k|K)=i\frac{1}{bln(1-q^2/\Lambda ^2)}[t^a_{(1)}][t^a_{(2)}]\gamma ^\mu _{(1)}
\gamma ^\nu _{(2)}D^{(0)}_{\mu \nu }(q),\quad q=p-k. \label{66}
\ee
Then for   the   ultraviolet   region   $-q^2\gg \Lambda ^2$   we   get
representation (\ref{63})  with  the  invariant  charge  (\ref{65}),
whereas for the infrared region $-q^2\ll \Lambda ^2$ we  obtain the behavior
which coincides  in  details   with   the   behavior   following   from
representation (\ref{61}) if we put
\[
g^2\kappa ^2\equiv \Lambda ^2/b.
\]
Besides, a more general parameterization of the form
\be
\tilde R(p;k|K)=i\frac{1}{bln(\xi -q^2/\Lambda ^2)}[t^a_{(1)}][t^a_{(2)}]\gamma ^\mu _{(1)}
\gamma ^\nu _{(2)}D^{(0)}_{\mu \nu }(q),\label{67}
\ee
can be considered. Here $\xi $  is  some  phenomenological  parameter
obeying the condition that $\xi \geq 1$. We  shall  call  parameterization
(\ref{67}) the generalized Richardson's parameterization meaning that
it leads to  the  standard  parameterization  (\ref{66})  at  $\xi =1$.
Remember, the function $D^{(0)}_{\mu \nu }$ entering in  the  R.H.S.  of
equality (\ref{67}) is a free gluon propagator, for which we shall
use the standard expression in the invariant gauge
\be
D_{\mu \nu }^{(0)}(q) = \frac{-1}{q^2+i\varepsilon }(g_{\mu \nu }+(d-1)
\frac{q_{\mu }q_{\nu }}{q^2+i\varepsilon }). \label{68}
\ee

Our  further  calculations  will  be  made with account of the
generalized Richardson's parameterization and  by  the  scheme  we
adhered to in  previous Sections.  Using  representation
(\ref{67}) for the quark--antiquark interaction quasipotential we get
\[
V_{q_1\bar{q}_2}(nM\vert \tilde{p}_1\sigma _{i_1}\tilde{p}_2\sigma _{i_2};
\tilde{k}_2\lambda _{j_2}\tilde{k}_1\lambda _{j_1}) = n_0\delta ^3(\vec{P}_M
-\vec{K}_M)
\delta _{\sigma _{f_1}}^{\lambda _{f_1}}\delta _{\sigma _{f_2}}^{\lambda _{f_2}}
\sum_{a}[t_{(1)}^{a}]_{\sigma _{c_1}}^{\lambda _{c_1}}[t_{(2)}^{a}]
_{\lambda _{c_2}}^{\sigma _{c_2}} \nonumber\\
\]
\[
\times\left[\bar{u}(\vec{p}_1\sigma _1)\gamma ^\mu u(\vec{k}_1\lambda _1)\bar{v}(\vec{k}_2\lambda _2)
\gamma _\mu v(\vec{p}_2\sigma _2)A(nM\vert \tilde{p}_1\tilde{p}_2;\tilde{k}_1
\tilde{k}_2)\right.+ \nonumber\\
\]
\be
+\left.(d-1)\bar{u}(\vec{p}_1\sigma _1)(n\gamma )u(\vec{k}_1\lambda _1)\bar{v}(\vec{k}_2\lambda _2)
(n\gamma )v(\vec{p}_2\sigma _2)B(nM\vert \tilde{p}_1\tilde{p}_2;\tilde{k}_1
\tilde{k}_2)\right]. \label{69}
\ee
The functions $A$ and  $B$ in the R.H.S.  of  equality  (\ref{69})
are found with the help of the following integrals
\ber
\lefteqn{A(nM\vert \tilde{p}_1\tilde{p}_2;\tilde{k}_1
\tilde{k}_2) = }\nonumber\\
&=& \frac{1}{(2\pi )^5b}\int_{-\infty }^{\infty }d\alpha \int_{-\infty }^{\infty }d\beta
(\frac{1}{\Delta /2+\alpha -i\varepsilon }+\frac{1}{\Delta /2-\alpha -i\varepsilon })\times  \nonumber\\
&\times &(\frac{1}{\Delta '/2+\beta -i\varepsilon }+\frac{1}{\Delta '/2-\beta -i\varepsilon })
\frac{1}{[p-k-(\alpha -\beta )n]^2+i\varepsilon }\times  \nonumber\\
&\times &\frac{1}{ln[\xi -([p-k-(\alpha -\beta )n]^2+i\varepsilon )/\Lambda ^2]},\ \label{70}
\eer
\[
B(nM\vert \tilde{p}_1\tilde{p}_2;\tilde{k}_1
\tilde{k}_2) = \nonumber\\
\]
\[
= \frac{1}{(2\pi )^5b}\int_{-\infty }^{\infty }d\alpha \int_{-\infty }^{\infty }d\beta
(\frac{1}{\Delta /2+\alpha -i\varepsilon }+\frac{1}{\Delta /2-\alpha -i\varepsilon })\times  \nonumber\\
\]
\[
\times (\frac{1}{\Delta '/2+\beta -i\varepsilon }+\frac{1}{\Delta '/2-\beta -i\varepsilon })
[(\alpha -\beta )^2-(\Delta -\Delta ')^2/4]\times \nonumber\\
\]
\be
\times \frac{1}{([p-k-(\alpha -\beta )n]^2+i\varepsilon )^2
ln[\xi -([p-k-(\alpha -\beta )n]^2+i\varepsilon )/\Lambda ^2]}.\ \label{71}
\ee
In the expressions (\ref{70}) and (\ref{71}) we use the same  notations  for
the functions $A$ and  $B$ as in formula (\ref{6}). Our next Sections
will be devoted to investigating these functions.

\subsection {Analysis of the Function $A$}

The integral, specifying the function  $A$,  can  be  conveniently
transformed by introducing new variables $x=\alpha -\beta $ and
$X=\frac{1}{2}(\alpha +\beta )$ instead of  the  integration  variables
$\alpha $ and $\beta $. One can integrate over the variable $X$ with the help
of  the  residue theorem, after which the expression for $A$
reduces to the single integral
\be
A=\frac{2\pi i}{(2\pi )^5b}\cdot \frac{-1}{(\kappa _1-
\kappa _2)}\cdot I_A,\label{72}
\ee
with
\ber
I_A = \int^{\infty }_{-\infty }dx(\frac{1}{\delta +x-i\varepsilon }+\frac{1}{\delta -x-i\varepsilon })\times
\nonumber\\\times (\frac{1}{\kappa _1-x-i\varepsilon }+\frac{1}{x-\kappa _2-i\varepsilon })\frac{1}
{ln[(\kappa '_1-x-i\varepsilon )(x-\kappa '_2-i\varepsilon )/\Lambda ^2]}. \label{73}
\eer
Here the introduced notations are:
\[
\kappa _{1,2} = (np-nk) \pm  \sqrt{-(p-k)^2_{\perp }},\quad
\kappa '_{1,2} = (np-nk) \pm  \sqrt{\xi \Lambda ^2-(p-k)^2_{\perp }},\nonumber \\
\]
\be
(p-k)^2_{\perp } = (p-k)^2 - (np-nk)^2, \quad \delta \equiv \frac{1}{2}(\Delta +\Delta ')
\nonumber
\ee
so that the following equalities are valid:
\[
(p-k-xn)^2+i\varepsilon =(x-\kappa _1+i\varepsilon )(x-\kappa _2-i\varepsilon ), \nonumber \\
\]
\[
(p-k-xn)^2-\xi \Lambda ^2+i\varepsilon =(x-\kappa '_1+i\varepsilon )(x-\kappa '_2-i\varepsilon ).\nonumber
\]

A detailed investigation of analytic  structure  of  the  integral
(\ref{73}) is contained in the Appendix B of this paper. But here we
give the result for the integral in a particular  evolution  gauge
where the normal vector $n$  points  along
the total momentum of the system, and for the case of equal masses
of  the  quark  and   antiquark:   $m_1=m_2=m$,   $np=nk=0$.   The
calculation of  integral (\ref{73}) in arbitrary gauge in the  case
when the masses of quark and antiquark are not equal are available
in the Appendix B. So, in the given particular evolution gauge there
is:
\ber
A(nM\vert \tilde{p}_1\tilde{p}_2;\tilde{k}_1\tilde{k}_2) =
A(M\vert p_{\perp };k_{\perp }) = \frac{1}{(2\pi )^3b\Lambda ^2}\times \nonumber \\[2ex]
\times \left[\frac{1}{(q^2_\Lambda -\delta ^2_\Lambda )ln(\xi +q^2_\Lambda -\delta ^2_\Lambda )}-\frac{1}{ln\xi }
\cdot\frac{\delta _\Lambda }{q_\Lambda (q^2_\Lambda -\delta ^2_\Lambda )}\right.+\nonumber \\[2ex]
+\frac{\delta _\Lambda }{(\xi -1)\sqrt{\xi -1+q^2_\Lambda }(\xi -1+q^2_\Lambda -\delta ^2_\Lambda )}+\nonumber
\\[2ex]
+\left.2\delta _\Lambda \int^{\infty }_{\sqrt{\xi }}\frac{dy}{y}\frac{1}{[ln^2(y^2-\xi )+\pi
^2]\sqrt{y^2+
q^2_\Lambda }(y^2+q^2_\Lambda -\delta ^2_\Lambda )}\right],\label{74}
\eer
with the following notations introduced:
\[
q_\Lambda =q/\Lambda ,\quad \delta _{\Lambda }=\delta /\Lambda ,\quad  q=\sqrt{-(p-k)^2_{\perp }},
\]
\[
\delta \equiv \sqrt{m^2-p^2_{\perp }}+\sqrt{m^2-k^2_{\perp }}-M.
\]
The R.H.S. of Eq. (\ref{74}) contains the terms  singular  at
$\xi \to 1$. However, one may easily verify  that at  $\xi \to 1$ the limit
does exist and is equal to
\ber
A(M\vert p_{\perp };k_{\perp })\vert _{\xi =1} \equiv  A_{R}(M\vert p_{\perp };k_{\perp }) =
 \frac{1}{(2\pi )^3b\Lambda ^2}\times \nonumber \\[2ex]
\times \left[\frac{1}{(q^2_\Lambda -\delta ^2_\Lambda )ln(1+q^2_\Lambda -\delta ^2_\Lambda )}-
\frac{\delta _\Lambda }{2q_\Lambda (q^2_\Lambda -\delta ^2_\Lambda )}
-\frac{\delta _\Lambda }{2q^3_\Lambda (q^2_\Lambda -\delta ^2_\Lambda )}\right.-\nonumber\\[2ex]
-\frac{\delta _\Lambda }{q_\Lambda (q^2_\Lambda -\delta ^2_\Lambda )^2}
+\left.2\delta _\Lambda \int^{\infty }_{1}\frac{dy}{y}\frac{1}{[ln^2(y^2-1)+\pi ^2]\sqrt{y^2+
q^2_\Lambda }(y^2+q^2_\Lambda -\delta ^2_\Lambda )}\right].\label{75}
\eer
Having now the expression for the function $A$, we can  study  how
it behaves in the limit when $\Lambda ^2\to \infty $. Taking in the R.H.S. of  Eq.
(\ref{74}) $\xi =e$  and  tending  $\Lambda ^2\to \infty $,  we come to
\be
A(M|p_{\perp };k_{\perp })\vert _{\xi =e, \Lambda ^2\to \infty }=\frac{1}{(2\pi )^3b}\cdot
\frac{1}{q(q+\delta )}.\label{76}
\ee
The R.H.S. of (\ref{76}) coincides with  the  previously  obtained
expression  for  the   function   $A^{(0)}$,   provided   we   put
$g^2=b^{-1}$. After an analogous limiting transition in the
R.H.S.  of  equality (\ref{75}), we find that
\be
A(M|p_{\perp };k_{\perp })\vert _{\xi =1, \Lambda ^2\to \infty } =
\frac{\Lambda ^2}{(2\pi )^3b}\left[\frac{1}{2q^3(q+\delta )}+
\frac{1}{2q^2(q+\delta )^2}\right], \label{77}
\ee
which, in  its  turn,  fully  coincides  with  the  earlier  found
expression for the function $A^{(1)}$ at  $g^2\kappa ^2=\Lambda ^2b^{-1}$, the
latter  defining  the contribution  from  the  infrared
singularities  of   the   gluon propagator. These  two
results -- (\ref{76}) and (\ref{77})-- are easy to understand if
one  turns  to the original integral (\ref{70}) specifying the $A$
function.  The consideration of the limit  $\Lambda ^2\to \infty $  in  the
R.H.S.  of  equality (\ref{70}) at $\xi =e$ and $\xi =1$  will
obviously  bring  us  to  the integrals of the functions $A^{(0)}$
and $A^{(1)}$,  respectively. (These functions were already
calculated in previous Sections). Therefore, the limiting relations
(\ref{76}) and (\ref{77}) correlate our calculations.

\subsection{Analysis of the Function $B$}

As before, the integral (\ref{71}), specifying the  $B$  function,
can  be  conveniently  transformed  with  the  help  of  the   new
integration variables $x=\alpha -\beta $ and $X=\frac{1}{2}(\alpha +\beta )$ and via
the integration over the variable $X$. As a result, the
expression for the function $B$ turns out to be  nothing  but  the
difference of two single integrals
\be
B=B^{(1)}-\frac{(\Delta -\Delta ')^2}{4}B^{(2)},\label{78}
\ee
and besides,
\[
B^{(1)} = \frac{2\pi i}{(2\pi )^3b(\kappa _1-\kappa _2)^2}I^{(1)}_B,\quad
B^{(2)} = \frac{2\pi i}{(2\pi )^3b(\kappa _1-\kappa _2)^2}I^{(2)}_B,
\]
with
\ber
I^{(1)}_B = \int^{\infty }_{-\infty }dx(\frac{1}{\delta +x-i\varepsilon }+\frac{1}{\delta -x-i\varepsilon })\times
\nonumber\\\times (\frac{1}{\kappa _1-x-i\varepsilon }+\frac{1}{x-\kappa _2-i\varepsilon })^2\frac{x^2}
{ln[(\kappa '_1-x-i\varepsilon )(x-\kappa '_2-i\varepsilon )/\Lambda ^2]}, \label{79}
\eer
\ber
I^{(2)}_B = \int^{\infty }_{-\infty }dx(\frac{1}{\delta +x-i\varepsilon }+\frac{1}{\delta -x-i\varepsilon })\times
\nonumber\\\times (\frac{1}{\kappa _1-x-i\varepsilon }+\frac{1}{x-\kappa _2-i\varepsilon })^2\frac{1}
{ln[(\kappa '_1-x-i\varepsilon )(x-\kappa '_2-i\varepsilon )/\Lambda ^2]}. \label{80}
\eer
Here we use again the notations from the expression for the function
$A$. The results below are the calculations of integrals
(\ref{79}) and (\ref{80}) in a particular evolution gauge and  for
equal  quark  and  antiquark  masses.  Again,   as   before,   the
complete exploration  of  these  integrals  is  presented  in  the
Appendix B. For the function $B^{(1)}$, therefore, we have
\ber
\lefteqn{B^{(1)}(M\vert p_{\perp };k_{\perp }) =}\nonumber\\
&=& -\frac{1}{(2\pi )^3b\Lambda ^2}\cdot
\frac{\delta _\Lambda }{2q_\Lambda } \left[
\frac{2\delta _\Lambda }{(q_\Lambda +\delta _\Lambda )(q^2_\Lambda -\delta ^2_\Lambda )ln(\xi +q^2_\Lambda -\delta ^2_\Lambda )}\right.-
\nonumber\\[2ex]
&-&\frac{1}{ln\xi }\cdot\frac{1}{(q^2_\Lambda -\delta ^2_\Lambda )}+
\frac{2\sqrt{\xi -1+q^2_\Lambda }}{(\xi -1)(\xi -1+q^2_\Lambda -\delta ^2_\Lambda )(q_\Lambda +
\sqrt{\xi -1+q^2_\Lambda })}+ \nonumber \\[2ex]
&+&\left.\int^{\infty }_{\sqrt{\xi }}\frac{dy}{y}\frac{4\sqrt{y^2+q^2_\Lambda }}{[ln^
2(y^2-\xi )+\pi ^2]
(q_\Lambda +\sqrt{y^2+q^2_\Lambda })(y^2+q^2_\Lambda -\delta ^2_\Lambda )}\right].\label{81}
\eer
Here, again, we have the terms singular at $\xi \to 1$. It makes  no
difficulty to verify that, as before, the limit at  $\xi \to 1$  does  exist
and looks like
\ber
B^{(1)}(M\vert p_{\perp };k_{\perp })\vert _{\xi =1} \equiv  B^{(1)}_{R}(M\vert p_{\perp };k_{\perp }) =
\frac{1}{(2\pi )^3b\Lambda ^2}\cdot
\frac{\delta _\Lambda }{2q_\Lambda }\left[\frac{1}
{2(q^2_\Lambda -\delta ^2_\Lambda )}\right.+\nonumber \\[2ex]
+\frac{1}{(q^2_\Lambda -\delta ^2_\Lambda )^2}-\frac{1}{4q^2_\Lambda (q^2_\Lambda -\delta ^2_\Lambda )}-
\frac{2\delta _\Lambda }{(q_\Lambda +\delta _\Lambda )(q^2_\Lambda -\delta ^2_\Lambda )ln(1+q^2_\Lambda -\delta ^2_\Lambda )}-
\nonumber \\[2ex]
-\left.\int^{\infty }_{1}\frac{dy}{y}\frac{4\sqrt{y^2+q^2_\Lambda }}{[ln^2(y^2-1)+\pi ^2]
(q_\Lambda +\sqrt{y^2+q^2_\Lambda })(y^2+q^2_\Lambda -\delta ^2_\Lambda )}\right] . \label{82}
\eer
Calculating the  integral  for  the  function  $B^{(2)}$  (in  the
above-- mentioned particular gauge), we are brought to :
\[
B^{(2)}(M\vert p_{\perp };k_{\perp }) = -\frac{1}{(2\pi )^3b\Lambda ^4}\cdot
\frac{1}{2q_\Lambda }\times \nonumber\\
\]
\[
\times \left[\frac{2}{(q_\Lambda +\delta _\Lambda )(q^2_\Lambda -\delta ^2_\Lambda )ln(\xi +q^2_\Lambda -\delta ^2_\Lambda )}-
\frac{1}{ln\xi }\cdot\frac{\delta _\Lambda }{q^2_\Lambda (q^2_\Lambda -\delta ^2_\Lambda )}\right.+\nonumber\\
\]
\[
+\frac{2\delta _\Lambda }{(\xi -1)\sqrt{\xi -1+q^2_\Lambda }(\xi -1+q^2_\Lambda -\delta ^2_\Lambda )(q_\Lambda +
\sqrt{\xi -1+q^2_\Lambda })}+ \nonumber \\
\]
\be
+\left.\int^{\infty }_{\sqrt{\xi }}\frac{dy}{y}\frac{4\delta _\Lambda }{[ln^2(y^2-\xi )+\pi ^2]
\sqrt{y^2+q^2_\Lambda }(q_\Lambda +\sqrt{y^2+q^2_\Lambda })(y^2+q^2_\Lambda -\delta ^2_\Lambda )}\right].
\label{83}
\ee
After the limiting transition $\xi \to 1$  in  the  R.H.S.  of  equality
(\ref{83}), the result will be
\ber
B^{(2)}(M\vert p_{\perp };k_{\perp })\vert _{\xi =1} \equiv  B^{(2)}_{R}(M\vert p_{\perp };k_{\perp }) =
\frac{1}{(2\pi )^3b\Lambda ^4}\cdot
\frac{1}{2q_\Lambda }
\left[\frac{\delta _\Lambda }{q^2_\Lambda (q^2_\Lambda -\delta ^2_\Lambda )^2}\right.+\nonumber \\
+\frac{\delta _\Lambda }{2q^2_\Lambda (q^2_\Lambda -\delta ^2_\Lambda )}+\frac{3\delta _\Lambda }{4q^4_\Lambda (q^2_\Lambda -\delta ^2_\Lambda )}
-\frac{2}{(q_\Lambda +\delta _\Lambda )(q^2_\Lambda -\delta ^2_\Lambda )ln(1+q^2_\Lambda -\delta ^2_\Lambda )}-\nonumber \\
-\left.\int^{\infty }_{1}\frac{dy}{y}\frac{4\delta _\Lambda }{[ln^2(y^2-1)+\pi ^2]
\sqrt{y^2+q^2_\Lambda }(q_\Lambda +\sqrt{y^2+q^2_\Lambda })(y^2+q^2_\Lambda -\delta ^2_\Lambda )}\right].
\label{84}
\eer
Expressions   (\ref{78}),   (\ref{81}), (\ref{83})   describe   the
analytic structure of the function $B$. It would  also
be useful to look at the behavior of this function  in  the  limit
$\Lambda ^2\to \infty $. Assuming  for  the  R.H.S.  of  (\ref{81})
$\xi =e$  and  tending
$\Lambda ^2\to \infty $, for the function $B^{(1)}$ we get
\be
B^{(1)}(M|p_{\perp };k_{\perp })\vert _{\xi =e, \Lambda ^2\to \infty }=\frac{1}{(2\pi )^3b}\cdot
\frac{\delta }{2q(q+\delta )^2}.\label{85}
\ee
After performing an analogous procedure in the R.H.S. of (\ref{83}),
for the function $B^{(2)}$ we find that
\be
B^{(2)}(M|p_{\perp };k_{\perp })\vert _{\xi =e, \Lambda ^2\to \infty }=-\frac{1}{(2\pi )^3b}\cdot
\frac{1}{2q}\left[\frac{1}{q^2(q+\delta )}+\frac{1}{q(q+\delta )^2}\right],
\label{86}
\ee
and, hence, the  complete  function  $B$  in  this  limiting  case
becomes
\ber
B(M|p_{\perp };k_{\perp })\vert _{\xi =e, \Lambda ^2\to \infty }=\frac{1}{(2\pi )^3b}\left[
\frac{\delta }{2q(q+\delta )^2}\right.+\nonumber \\[2ex]
+\left.\frac{(\sqrt{m^2-p^2_{\perp }}-\sqrt{m^2-k^2_{\perp }})^2}{2q}\left(
\frac{1}{q^2(q+\delta )}+\frac{1}{q(q+\delta )^2}\right)\right].\label{87}
\eer
Remember,   here   $\delta =\sqrt{m^2-p^2_\perp }+\sqrt{m^2-k^2_\perp }-M$,    and
besides, we take into account that $\frac{1}{4}(\Delta -\Delta ')^2=
(\sqrt{m^2-p^2_\perp }-\sqrt{m^2-k^2_\perp })^2$.   Expression    (\ref{87})
coincides exactly with the expression for the function
$B^{(0)}$ that was obtained before under the condition that
$g^2=b^{-1}$.

The limiting transition $\Lambda ^2\to \infty $ in the R.H.S.  of  Eq.
(\ref{82}) for the function $B^{(1)}$ leads to
\be
B^{(1)}(M|p_{\perp };k_{\perp })\vert _{\xi =1, \Lambda ^2\to \infty }=\frac{\Lambda ^2}{(2\pi )^3b}\cdot
\frac{1}{4q} \left[\frac{3\delta }{2q^2(q+\delta )^2}-\frac{\delta ^2}
{q^2(q+\delta )^3}\right].\label{88}
\ee
Repeating the same limiting procedure in the R.H.S. of Eq. (84),
for the function $B^{(2)}$ we obtain that
\ber
B^{(2)}(M|p_{\perp };k_{\perp })\vert _{\xi =1, \Lambda ^2\to \infty }=-\frac{\Lambda ^2}{(2\pi )^3b}\cdot
\frac{1}{4q}\times \nonumber \\[2ex]
\times \left[\frac{1}{q^2(q+\delta )^3}+\frac{3}{2q^3(q+\delta )^2}+\frac{3}{2q^4(q+\delta )}
\right].\label{89}
\eer
Thus, in the given limiting case for the complete function $B$  we
have
\ber
B(M|p_{\perp };k_{\perp })\vert _{\xi =1, \Lambda ^2\to \infty }=\frac{\Lambda ^2}{(2\pi )^3b}\cdot
\frac{1}{4q^3} \left[ \frac{3\delta }{2(q+\delta )^2}-\frac{\delta ^2}
{(q+\delta )^3}\right.+\nonumber \\[2ex]+\left.(\sqrt{m^2-p^2_{\perp }}-
\sqrt{m^2-k^2_{\perp }})^2
\left(\frac{1}{(q+\delta )^3}+\frac{3}{2q(q+\delta )^2}+\frac{3}{2q^2(q+\delta )}
\right)\right].\label{90}
\eer
The resulting expression (\ref{90}) for the function $B$ coincides
with the expression for the function $B^{(1)}$ from  Eq. (\ref{23}) at
$g^2\kappa ^2=\Lambda ^2b^{-1}$. Looking back at the original
integrals, (\ref{79}) and (\ref{80}), specifying the function  $B$,
we see that the results  (\ref{87})  and  (\ref{90}),  similar  to
those of the previous Section,  have  not  come  unexpected  but
rather as a correlation of our calculations.
Further, it would be convenient to explore the properties  of  the
dynamic functions, $A$ and $B$,  going  over  to  the  configuration
space; that is what we'll do in the next Section.

\subsection{Analysis of the Dynamic Functions in Configuration
Space}

To analyze the dynamic functions $A$ and $B$ in the  configuration
space, pass over in these functions to new variables defined as
\[
{\tilde{p}}_i = L(n) {\ovc{\tilde{p}}}_i,\qquad
{\tilde{k}}_i = L(n) {\ovc{\tilde{k}}}_i,
\]
with $L(n)$  the matrix of the Lorentz boost which has the
property
\[
L^{-1}(n)n=(1,\vec 0).
\]
It is easy to see  that  in  the  given  particular  gauge the new
momentum variables  are as follows:
\[
{\ovc{p}}_{1\perp } = -{\ovc{p}}_{2\perp } = {\ovc{p}}_{\perp } = (0,\ovc{\vec{p}}),
\qquad {\ovc{k}}_{1\perp } = -{\ovc{k}}_{2\perp } = {\ovc{k}}_{\perp } = (0,\ovc{\vec{k}}),
\]
\[
\ovc{\vec{p}} = {\ovc{\vec{p}}}_1 = -{\ovc{\vec{p}}}_2,\qquad
\ovc{\vec{k}} = {\ovc{\vec{k}}}_1 = -{\ovc{\vec{k}}}_2.
\]
Besides,
\[
p_{\perp }^2 = ({\ovc{p}}_{\perp })^2 = -(\ovc{\vec{p}})^2,\qquad
k_{\perp }^2 = ({\ovc{k}}_{\perp })^2 = -(\ovc{\vec{k}})^2,\nonumber
\]
and hence,
\[
q=\sqrt{-(p-k)^2_{\perp }}=\sqrt{(\ovc{\vec{p}}-\ovc{\vec{k}})^2}=
\vert \ovc{\vec{p}}-\ovc{\vec{k}}\vert ,
\]
\[
\delta =\sqrt{m^2+\ovc{{\vec{p}}\,^2}}+\sqrt{m^2+\ovc{{\vec{k}}\,^2}}-M.
\]
The dynamic functions $A$ and $B$, as their  explicit  expressions
show, are nonlocal functions depending on the  spectral  parameter
$M$, the latter taking  the  values  of  the  total  energy  of  a
two--fermion  system.  This  result   follows   from   a   thorough
consideration  of  the  problem   of   a   relativistic   particle
interaction in the framework of the local  quantum  field  theory.
Pay attention to that the whole nonlocality of the dynamic  functions
concentrates in  the  quantity  $\delta $  contained  therein.  The  $\delta $
determines the off energy  shell  continuation  symmetric  in  the
particle momenta of the  initial  and  final  states.  On  energy
shell, $\delta $ and, consequently, the $B$ functions turn into zero. As
for the $A$  function,  on  energy  shell  it  becomes  local  and
coincides with the known Richardson's potential.

As we already know, to achieve the locality of the  dynamic  functions
by restricting them to  the energy  shell means  rather a
destructive trick, since in this case many  of  the  dynamic
properties of a relativistic interaction are lost.  Another
way to locally approximate the dynamic functions is to  keep  to  only
such  configurations  of  interacting  particles  for  which   the
conditions $\ovc{\vec{p}\,^2}/m^2\ll 1,  \ovc{\vec{k}\,^2}/m^2\ll 1$ are
true. Then the quantity $\delta $ is supposed to be equal to the  defect
of the mass of the system or, which is the same,  to the binding
energy with the opposite sign,  whereas  the dynamic   functions
determining  the  potential  become  local  functions   with   the
dependence on the binding energy  of  the  system.  After  such  a
procedure, for the function $A_R$,  for  instance,  the  following
local approximation will be obtained:
\[
A_R(M\vert p_{\perp };k_{\perp }) \cong  A_{R}(\varepsilon ,\Lambda ;q) =
\]
\[
=\frac{G}{\Lambda ^2}\left[\frac{1}{[q^2_\Lambda -(\varepsilon _\Lambda +i0)^2]
ln[1+q^2_\Lambda -(\varepsilon _\Lambda +i0)^2]}\right.+\nonumber \\
\]
\[
+\frac{\varepsilon _\Lambda }{2q_\Lambda [q^2_\Lambda -(\varepsilon _\Lambda +i0)^2]}
+\frac{\varepsilon _\Lambda }{2q^3_\Lambda [q^2_\Lambda -(\varepsilon _\Lambda +i0)^2]}
+\frac{\varepsilon _\Lambda }{q_\Lambda [q^2_\Lambda -(\varepsilon _\Lambda +i0)^2]^2}-\nonumber \\
\]
\be
-\left.2\varepsilon _\Lambda \int^{\infty }_{1}\frac{dy}{y}\frac{1}{[ln^2(y^2-1)+\pi ^2]\sqrt{y^2+
q^2_\Lambda }[y^2+q^2_\Lambda -(\varepsilon _\Lambda +i0)^2]}\right].\label{91}
\ee
Here $G=[(2\pi )^3b]^{-1}, \varepsilon _\Lambda \equiv \varepsilon /\Lambda , \varepsilon =M-2m$ is the binding energy  of
the system, and there in the R.H.S. of Eq. (\ref{91}) an  explicit
indication is contained by which rule to bypass the  singularities
in the integral, determining the  transition to  the  configuration
space.
\be
\tilde A_R(\varepsilon ,\Lambda ;r)=\frac{4\pi }{r}
\int^{\infty }_{0}qdq sin(qr)A_R(\varepsilon ,\Lambda ;q).\label{92}
\ee
Note, this prescription is, on the one hand, due  to  the  causal
structure of the local quantum field theory, which is used as  the
framework  for  the  construction  of   the   given   single--time
formalism, and, on the other hand, it itself guarantees the  causal
properties of the dynamic equations in the  single--time
formalism [8,17].

Expression (\ref{91}) for  the  function  $A_R$  has  five  terms.
Hence, in order to pass over to the configuration space,  one  has
to calculate, respectively, five  integrals  of  form  (\ref{92}).
Each of the five integrals can be found in the Appendix~C.
Here we only present the complete result for  the  whole  function
$A_R$  in  the  particular  evolution  gauge.  As   usual,   we'll
distinguish between the regions of positive and negative values of
the binding energy and, besides, restrict ourselves to the  values
of the binding energy which fulfil the condition
$\varepsilon ^2_\Lambda <1$.  Thus,
at a positive binding energy, for the function $A_R$ we get
\ber
\tilde A_R(\varepsilon .\Lambda ;r)=\frac{G\pi ^2}{r}\left[2e^{i\varepsilon r}(1+\frac{i}{\varepsilon _\Lambda }\Lambda r)-\frac{2}
{\pi }[a(\varepsilon r)-\frac{1}{\varepsilon _\Lambda }\Lambda rb(\varepsilon r)]\right.-\nonumber \\[2ex]
-\frac{\Lambda r}{\varepsilon _\Lambda }[i+\frac{2}{\pi }b(\mu r)]\vert _{\mu \to 0}-4\int^{\infty }_{1}\frac{dy}
{y}\frac{exp(-\Lambda r\sqrt{y^2-\varepsilon ^2_\Lambda })}{ln^2(y^2-1)+\pi ^2}-\nonumber
\\[2ex]
-\left.\frac{8\varepsilon _\Lambda }{\pi }\int^{\infty }_{1}\frac{dy}{y}\frac{1}{ln^2(y^2-1)+
\pi ^2}\int^{\infty }_{0}tdt\frac{sin(\Lambda rt)}{\sqrt{t^2+y^2}(t^2+y^2-\varepsilon ^2_\Lambda )}
\right], \label{93}
\eer

In the region of negative values of the  binding  energy
$\varepsilon \equiv -\bar
\varepsilon <0$, for the function $A_R$ we find
\ber
\tilde A_R(\varepsilon ,\Lambda ;r)=\frac{G\pi ^2}{r}\left[\frac{2}{\pi }[a(\bar{\varepsilon }r)-
\frac{1}{\bar{\varepsilon }_\Lambda }\Lambda rb(\bar{\varepsilon }r)]\right.+\nonumber \\[2ex]
+\frac{\Lambda r}{\bar{\varepsilon }_\Lambda }[i+\frac{2}{\pi }b(\mu r)]\vert _{\mu \to 0}-4\int^{\infty }_{1}\frac{dy}
{y}\frac{exp(-\Lambda r\sqrt{y^2-\bar{\varepsilon }^2_\Lambda })}
{ln^2(y^2-1)+\pi ^2}+\nonumber \\[2ex]
+\left.\frac{8\bar{\varepsilon }_\Lambda }{\pi }\int^{\infty }_{1}\frac{dy}{y}\frac{1}{ln^2(y^2-1)+
\pi ^2}\int^{\infty }_{0}tdt\frac{sin(\Lambda rt)}{\sqrt{t^2+y^2}(t^2+y^2-\bar{\varepsilon }^2_\Lambda )}
\right]. \label{94}
\eer
When calculating the integral (\ref{92}) for the  function  $A_R$,
we come across the divergent integral stemming from the third term
in the R.H.S. of  Eq.  (\ref{91}).  We  work  with  this  integral
exploiting the standard regularization procedure. This fact  shows
itself in the presence of  an  infinite  constant  $b(0)$  in  the
R.H.S.'s of Eqs. (\ref{93}) and (\ref{94}). Below, this fact  will
be further discussed.

When the local approximation for the function $B_R$ is constructed
in that way, the function $B^{(2)}_R$ will yield no  contribution,
since in the local limit under  consideration  the  factor  before
this function,
$(\sqrt{m^2+\ovc{\vec{p}\,^2}}-\sqrt{m^2+\ovc{\vec{k}\,^2}})^2$,
turns into zero. As a result,  the  local  approximation  for  the
function $B_R$ will be:
\[
B_R(M\vert p_{\perp };k_{\perp }) \cong  B^{(1)}_{R}(\varepsilon ,\Lambda ;q) =\nonumber \\
\]
\ber
=\frac{G}{\Lambda ^2}\cdot
\frac{\varepsilon _\Lambda }{2q_\Lambda }\left[
\frac{-2\varepsilon _\Lambda }{(q_\Lambda -\varepsilon _\Lambda -i0)[q^2_\Lambda -(\varepsilon _\Lambda +io)^2]ln[1+q^2_\Lambda -(\varepsilon _\Lambda +i0)^2]}
\right.-\nonumber \\[2ex]
-\frac{1}{2[q^2_\Lambda -(\varepsilon _\Lambda +i0)^2]}-\frac{1}{[q^2_\Lambda -(\varepsilon _\Lambda +i0)^2]^2}
+\frac{1}{4q^2_\Lambda [q^2_\Lambda -(\varepsilon _\Lambda +i0)^2]}+\nonumber \\[2ex]
+\left.\int^{\infty }_{1}\frac{dy}{y}\frac{4\sqrt{y^2+q^2_\Lambda }}{[ln^2(y^2-1)+\pi ^2]
(q_\Lambda +\sqrt{y^2+q^2_\Lambda })[y^2+q^2_\Lambda -(\varepsilon _\Lambda +i0)^2]}\right]. \label{95}
\eer
Here, again, to go over to the configuration  space,  one  has  to
calculate five integrals of form (\ref{92}), corresponding to  the
five terms entering in the R.H.S. of Eq. (\ref{95}). Referring  the
reader, as in the previous case with the function  $A_R$,  to  the
Appendix~C where the details of calculations can be found, here  we
write down the complete result  for  the  function  $B_R$  in  the
configuration  space  in  the  particular  evolution  gauge.   For
positive values of the binding energy we have that
\ber
\tilde B_R(\varepsilon ,\Lambda ;r)=-\frac{G\pi ^2}{r}\left[e^{i\varepsilon r}[i\varepsilon r-\frac{i}{2\varepsilon _\Lambda }\Lambda r-
\frac{1}{2}(\Lambda r)^2-\frac{1}{3}\varepsilon ^2_\Lambda ]\right.+\nonumber \\[2ex]
+\frac{2}{\pi }[(\frac{3}{4\varepsilon ^2_\Lambda }-\frac{1}{2})a(\varepsilon r)+\frac{1}{2\varepsilon _\Lambda }\Lambda r
b(\varepsilon r)]+\frac{\Lambda r}{4\varepsilon _\Lambda }[i+\frac{2}{\pi }b(\mu r)]\vert _{\mu \to 0}+\nonumber
\\[2ex]
+\frac{4\varepsilon ^2_\Lambda }{\pi }\int^{\infty }_{0}dt\frac{exp(-\Lambda rt)[\varepsilon _\Lambda ln\vert 1-\varepsilon ^2_\Lambda -t^2\vert
+\pi t\Theta (t^2+\varepsilon ^2_\Lambda -1)]}{(t^2+\varepsilon ^2_\Lambda )^2[ln^2\vert 1-\varepsilon ^2_\Lambda -t^2\vert +
\pi ^2\Theta (t^2+\varepsilon ^2_\Lambda -1)]}-\nonumber\\[2ex]
-\left.\frac{8\varepsilon _\Lambda }{\pi }\int^{\infty }_{1}\frac{dy}{y}\frac{1}{ln^2(y^2-1)+\pi ^2}
\int^{\infty }_{0}dt\frac{\sqrt{t^2+y^2}sin(\Lambda rt)}{(t+\sqrt{t^2+y^2})
(t^2+y^2-\varepsilon ^2_\Lambda )}\right]. \label{96}
\eer
At negative values of the binding energy, $\varepsilon =-\bar \varepsilon <o$, $\bar \varepsilon >0$,
for the function $B_R$ there is
\[
\tilde B_R(\varepsilon ,\Lambda ;r)=\nonumber\\
\]
\[
=-\frac{G\pi ^2}{r}\left[\frac{2}{\pi }[(\frac{1}{2}-
\frac{3}
{4\bar{\varepsilon }^2_\Lambda })a(\bar{\varepsilon }r)-\frac{1}{2\bar{\varepsilon }_\Lambda }\Lambda rb(\bar{\varepsilon }r)]-
\frac{\Lambda r}{4\bar{\varepsilon }_\Lambda }[i+\frac{2}{\pi }b(\mu r)]\vert _{\mu \to 0}\right.+\nonumber\\
\]
\[
+\frac{4\bar{\varepsilon }^2_\Lambda }{\pi }\int^{\infty }_{0}dt\frac{exp(-\Lambda rt)
[-\bar{\varepsilon }_\Lambda ln\vert 1-\bar{\varepsilon }^2_\Lambda -t^2\vert
+\pi t\Theta (t^2+\bar{\varepsilon }^2_\Lambda -1)]}{(t^2+\bar{\varepsilon }^2_\Lambda )^2[ln^2\vert 1-\bar{\varepsilon }^2_\Lambda -t^2\vert +
\pi ^2\Theta (t^2+\bar{\varepsilon }^2_\Lambda -1)]}+\nonumber \\
\]
\be
+\left.\frac{8\bar{\varepsilon }_\Lambda }{\pi }\int^{\infty }_{1}\frac{dy}{y}
\frac{1}{ln^2(y^2-1)+\pi ^2}
\int^{\infty }_{0}dt\frac{\sqrt{t^2+y^2}sin(\Lambda rt)}{(t+\sqrt{t^2+y^2})
(t^2+y^2-\bar{\varepsilon }^2_\Lambda )}\right]. \label{97}
\ee
In expressions (\ref{96}) and (\ref{97}) for the  function  $B_R$,
the  same  infinite  constant  $b(0)$  occurs  as  in  expressions
(\ref{93}) and (\ref{94})  describing  the  function  $A_R$.  This
infiniteness is due to the  divergency  of  the  integral  arising
from the integration of the fourth term  in  the  R.H.S.  of  Eq.
(\ref{95})  when  going  over  to  the  configuration  space.  The
situation with  the  divergencies  resembles  the  case  when  the
contribution from the  infrared  singularities  of  gluon  Green's
functions  into   the   two--quark   interaction   potential   was
investigated.  The  analogy  will  seem  still  deeper  if  we
remember that there exists the special gauge $d=-3$,  which  leads
to the cancellation of the divergencies in the linear combination
\[
V=A+(d-1)B,\nonumber
\]
determining  the  spin  independent  part   of   the   interaction
potential. So, here is the final result for the function
\[
V_A=V\vert _{d=-3}=A-4B.\nonumber
\]
Introducing an analogous linear combination:
\be
V_R=A_R+(d-1)B_R,\label{98}
\ee
one can easily get convinced that in the case under  consideration
in the gauge $d=-3$ the above divergencies cancel out, and we  get
the final function
\be
V_{AR}=V_R\vert _{d=-3}=A_R-4B_R.\label{99}
\ee
In the region  of  positive  values  of  the  binding  energy  the
expression for the function $V_{AR}$ becomes
\ber
V_{AR}(\varepsilon ,\Lambda ;r)=\frac{G\pi ^2}{r}\left\{2e^{i\varepsilon r}\left[1-\frac{2}{3}\varepsilon ^2_\Lambda +
2i\varepsilon r-(\Lambda r)^2\right]\right.+\nonumber \\[2ex]
+\frac{6}{\pi }\left[(\frac{1}{\varepsilon ^2_\Lambda }-1)a(\varepsilon r)+\frac{1}{\varepsilon ^2_\Lambda }\,\varepsilon rb(\varepsilon r)
\right]-4\int^{\infty }_{1}\frac{dy}{y}\frac{exp(-\Lambda r\sqrt{y^2-\varepsilon ^2_\Lambda })}
{ln^2(y^2-1)+\pi ^2}-\nonumber \\[2ex]
-\frac{8\varepsilon _\Lambda }{\pi }\int^{\infty }_{1}\frac{dy}{y}\frac{1}{ln^2(y^2-1)+\pi ^2}
\int^{\infty }_{0}tdt\frac{sin(\Lambda rt)}{\sqrt{t^2+y^2}
(t^2+y^2-\varepsilon ^2_\Lambda )}+\nonumber \\[2ex]
+\frac{16\varepsilon ^2_\Lambda }{\pi }\int^{\infty }_{0}dt\frac{exp(-\Lambda rt)[\varepsilon _\Lambda ln\vert 1-\varepsilon ^2_\Lambda -t^2\vert
+\pi t\Theta (t^2+\varepsilon ^2_\Lambda -1)]}{(t^2+\varepsilon ^2_\Lambda )^2[ln^2\vert 1-\varepsilon ^2_\Lambda -t^2\vert +
\pi ^2\Theta (t^2+\varepsilon ^2_\Lambda -1)]}-\nonumber \\[2ex]
-\left.\frac{32\varepsilon _\Lambda }{\pi }\int^{\infty }_{1}\frac{dy}{y}
\frac{1}{ln^2(y^2-1)+\pi ^2}
\int^{\infty }_{0}dt\frac{\sqrt{t^2+y^2}sin(\Lambda rt)}{(t+\sqrt{t^2+y^2})
(t^2+y^2-\varepsilon ^2_\Lambda )}\right\}.\label{100}
\eer
At negative  values  of  the  binding  energy  for  the  function
$V_{AR}$ we find
\[
V_{AR}(\varepsilon ,\Lambda ;r)= \nonumber \\
\]
\[
=\frac{G\pi ^2}{r}\left\{
\frac{6}{\pi }\left[(1-\frac{1}{\bar{\varepsilon }^2_\Lambda })a(\bar{\varepsilon }r)-
\frac{1}{\bar{\varepsilon }^2_\Lambda }\bar{\varepsilon }rb(\bar{\varepsilon }r)
\right]-4\int^{\infty }_{1}\frac{dy}{y}\frac{exp(-\Lambda r\sqrt{y^2-
\bar{\varepsilon }^2_\Lambda })}
{ln^2(y^2-1)+\pi ^2}\right.+\nonumber \\
\]
\[
+\frac{8\bar{\varepsilon }_\Lambda }{\pi }\int^{\infty }_{1}\frac{dy}{y}\frac{1}{ln^2(y^2-1)+\pi ^2}
\int^{\infty }_{0}tdt\frac{sin(\Lambda rt)}{\sqrt{t^2+y^2}
(t^2+y^2-\bar{\varepsilon }^2_\Lambda )}+\nonumber \\
\]
\[
+\frac{16\bar{\varepsilon }^2_\Lambda }{\pi }\int^{\infty }_{0}dt\frac{exp(-\Lambda rt)[-\bar \varepsilon _\Lambda ln\vert 1-
\bar{\varepsilon }^2_\Lambda -t^2\vert
+\pi t\Theta (t^2+\bar{\varepsilon }^2_\Lambda -1)]}{(t^2+\bar{\varepsilon }^2_\Lambda )^2[ln^2\vert 1-\bar{\varepsilon }^2_\Lambda -t^2\vert +
\pi ^2\Theta (t^2+\bar{\varepsilon }^2_\Lambda -1)]}+\nonumber \\
\]
\be
+\left.\frac{32\bar{\varepsilon }_\Lambda }{\pi }\int^{\infty }_{1}\frac{dy}{y}
\frac{1}{ln^2(y^2-1)+\pi ^2}
\int^{\infty }_{0}dt\frac{\sqrt{t^2+y^2}sin(\Lambda rt)}{(t+\sqrt{t^2+y^2})
(t^2+y^2-\bar{\varepsilon }^2_\Lambda )}\right\}.\label{101}
\ee
It may be proved that the known Richardson's  potential  is  derived
from the function $V_{AR}$  in  the  zero  binding  energy  limit.
Passing over to the limit of zero binding energy, in  any  of
the  expressions,  (\ref{100})  or  (\ref{101}),  for  the  function
$V_{AR}$ we get, as should really be, one and the same  result  of
the form
\be
\lim_{\varepsilon \to 0}V_{AR}(\varepsilon ,\Lambda ;r) =
G\pi ^2\left[\frac{f(\Lambda r)}{r}-\Lambda ^2r+C\right],\label{102}
\ee
where
\[
f(x)\equiv 1-4\int^{\infty }_{1}\frac{dt}{t}\frac{exp(-xt)}{ln^2(t^2-1)+\pi ^2},
\]
\[
C\equiv \lim_{\varepsilon \to 0}\frac{2\Lambda ^2}{\pi \varepsilon }.
\]
The comparison of Eqs. (\ref{100}), (\ref{101}) and (\ref{102}) shows
that the energy dependence changes  considerably  the
nature of the interaction: the smooth  behavior  of  the  function
$V_{AR}$ in the region of negative values of  the  binding  energy
goes into  oscillations  at  its  positive  values.  Moreover,  we
observe that in the region of large distances  the
function $V_{AR}$ starts to behave essentially in a different way depending
on which particular values the binding energy takes. For  positive
values of the binding energy formula (\ref{100}) gives
\be
V_{AR}(\varepsilon ,\Lambda ;r) = -2\pi ^2G\Lambda ^2re^{i\varepsilon r},
\quad r\gg \frac{1}{\varepsilon }>\frac{1}{\Lambda }.\label{103}
\ee
The asymptotic behavior of the function $V_{AR}$ in the region  of
large distances and for  negative  values  of  the  binding  energy
follows from expression (\ref{101}) and has the form
\be
V_{AR}(\varepsilon ,\Lambda ;r)=\frac{4\pi G\Lambda ^2}{\bar \varepsilon }\cdot \frac{1}{(\bar \varepsilon r)^2}
\left[1-4(1+\frac{\bar \varepsilon ^2_\Lambda }{ln(1-\bar
\varepsilon ^2_\Lambda )})+
\frac{3}{2}\bar \varepsilon ^2_\Lambda \right],\label{104}
\ee
\[
r\gg \frac{1}{\bar \varepsilon }>\frac{1}{\Lambda }.
\]
Therefore, at negative values of the binding energy  the  function
$V_{AR}$ is characterized at  large  distances  by  a rapider
than the Coulomb decrease, whereas at positive  values
of the binding energy there appear oscillations  of  the  function
$V_{AR}$, their amplitude growing linearly with distance.

In the region of small distances, $r\ll   \frac{1}{\Lambda }<\frac{1}{\vert \varepsilon \vert }$,
from  expressions  (\ref{100})  and  (\ref{101})  for  the  function
$V_{AR}$ we get

 a) the binding energy positive $\varepsilon >0$:
\be
V_{AR}(\varepsilon ,\Lambda ;r)=
\frac{G\pi ^2}{r}\left[c(\varepsilon _\Lambda )+\frac{1}
{ln(\frac{1}{\Lambda r}\sqrt{1+(\varepsilon r)^2})}\right],\label{105}
\ee

 b) the binding energy negative $\varepsilon =-\bar \varepsilon <0$:
\be
V_{AR}(\varepsilon ,\Lambda ;r)=
\frac{G\pi ^2}{r}\left[\bar c(\bar \varepsilon _\Lambda )+
\frac{1}{ln(\frac{1}{\Lambda r}\sqrt{1+
(\bar \varepsilon r)^2})}\right],\label{106}
\ee
the functions $c$ and $\bar c$ here being of the form:
\[
c(\varepsilon _\Lambda )=
-2-\frac{4}{3}\varepsilon ^2_\Lambda +\frac{3}{\varepsilon ^2_\Lambda }+
\]
\be
+\frac{16\varepsilon ^2_\Lambda }{\pi }\int^{\infty }_{0}dt\frac{\varepsilon _\Lambda ln\vert 1-\varepsilon ^2_\Lambda -t^2\vert +
\pi t\Theta (t^2+\varepsilon ^2_\Lambda -1)}{(t^2+\varepsilon ^2_\Lambda )^2[ln^2\vert 1-\varepsilon ^2_\Lambda -t^2\vert +\pi ^2\Theta (t^2+\varepsilon ^2_\Lambda -
1)]},\label{107}
\ee
\[
\bar{c}(\bar{\varepsilon }_\Lambda )=
2-\frac{3}{\bar{\varepsilon }^2_\Lambda }+
\]
\be
+\frac{16\bar{\varepsilon }^2_\Lambda }{\pi }\int^{\infty }_{0}dt\frac{-\bar{\varepsilon }_\Lambda ln\vert 1-
\bar{\varepsilon }^2_\Lambda -t^2\vert +
\pi t\Theta (t^2+\bar{\varepsilon }^2_\Lambda -1)}{(t^2+\bar{\varepsilon }^2_\Lambda )^2[ln^2\vert 1-
\bar{\varepsilon }^2_\Lambda -t^2\vert +\pi ^2\Theta (t^2+\bar{\varepsilon }^2_\Lambda -1)]}.\label{108}
\ee
Both the function $c$, and the function $\bar c$ may be  shown  to
vanish at zero binding energy limit. So, we  have   the
following property of these functions:
 \[
c(0)=\bar c(0)=0.
\]
Hence, it is only in the zero binding energy limit that we get  an
asymptotically free behavior of the function $V_{AR}$ , coinciding
with the behavior of the Richardson's potential at small  distances.
In  the  case  of  a  nonzero  binding   energy,   according to
expressions (\ref{105}) and (\ref{106}),  the  following  asymptotic
representation will be valid for the function  $V_{AR}$  at  small
distances:
\be
V_{AR}(\varepsilon ,\Lambda ;r)=\frac{\alpha (\varepsilon _\Lambda ,\Lambda r)}{r},\quad
r\ll \frac{1}{\Lambda }<\frac{1}{|\varepsilon |}.\label{109}
\ee
In this expression, the running coupling constant has an  explicit
energy dependence, and
\be
\alpha (\varepsilon _\Lambda ,\Lambda r)\vert _{\varepsilon =0}=
\alpha _R(\Lambda r)=\frac{G\pi ^2}{ln(\frac{1}{\Lambda r})}.\label{110}
\ee
Also,
\ber
\alpha (\varepsilon _\Lambda ,\Lambda r)\vert _{r=0}=G\pi ^2c(\varepsilon _\Lambda ),\quad \varepsilon >0,\nonumber \\
=G\pi ^2\bar{c}(\bar{\varepsilon }_\Lambda ),\quad \varepsilon =-\bar{\varepsilon }<0.\nonumber
\eer
The conclusion we are driven to is that when the binding energy is
different from zero, the function $V_{AR}$,  which  is  the  local
approximation of  the  two--quark  interaction  quasipotential  in
quantum chromodynamics, has a Coulomb singularity at zero.

\section{Conclusion}

In this paper the results of calculating the interaction
quasipotential for two quarks in QCD by early developed single-time
formalism in QFT for two-fermion systems have been presented. At
the first step we obtained analytical expressions for the
quark-antiquark interaction quasipotential in one-gluon exchange
approximation which explicitly take into account the structure
of the initial quantum field theory gauge model.

Then we have studied the influence of the infrared
singularities of the gluon Green`s functions on the behaviour of
the interaction potential for two quarks in QCD. The singular
behaviour of the gluon Green's function of the form $\kappa ^2/k^4$ is
known to be the result of nonperturbative investigations of the
infrared region in QCD. Therefore the results we obtained may also be
considered as going beyond the perturbative theory
when calculating quark--quark forces in the framework of the
fundamental QCD Lagrangian. Our consideration of the quark
interaction problem shows that the generally accepted notions,
that the singularity of the gluon propagator $\kappa ^2/k^4$ corresponds to
the interaction potential linearly growing with distance
are not quite correct. As can easily be seen from formulae
(\ref{57},\ref{59}) the indicated correspondence is restored only at zero
binding energy. Moreover, we have shown that when the binding
energy is negative the infrared singularity of the gluon
propagator does not lead to a potential linearly growing with
distance. However when going over to the region of positive
values for the binding energy there appear oscillations with the
amplitude linearly increasing with distance. The question what
such oscillations have to do with the confinement problem
remains open. At the same time one should note, that the
analogy with the solid state physics, where we find
oscillating potentials, allows us to consider such
oscillations of forces as a manifestation of the
quasicrystal structure of the vacuum in QFT.

As we have already noted, a consistent relativistic
consideration of the two body problem in the framework of local
QFT brings us to a nontrivial dependence of the interaction
potential on energy. Such energy dependence assigns the
interaction potential with rather unusual properties in the
configuration space which could hardly be imagined if one
stick to habitual quantum mechanical intuition. In particular,
the energy dependence of the
interaction potential results in the fact that the properties of
the forces qualitatively change during the transition from the
discrete spectrum to the region of the continuous one: a smooth
behaviour of the interaction potential in the discrete spectrum
is replaced by the oscillations in the continuous spectrum, and
this change in the behaviour of the forces is universal, i.e.
independent of concrete quantum field theory model and of used
approximations. The causal structure of local quantum field
theory displays in such manner. Recent studies [18] revealed
an extremely interesting property of the oscillating
potentials: such potentials lead to the appearance of discrete
levels in the continuous spectrum.

Here we were also interested in the problem of the gauge
dependence of the interaction potential. Here at least two
observed facts seem to be rather important. First, there is a
special gauge $d=-3$, where one manage correctly to
describe the infrared region in terms of the
interaction quasipotential. Only in this gauge the divergences are
canceled, and we arrive at a finite result for the interaction
quasipotential in the configuration space with an account of the
infrared singularities of gluon propagator. Secondly, a very
weak dependence of the interaction quasipotential on the gauge
parameter $d^{(0)}$ in the discrete spectrum becomes essentially
stronger when going over to the continuous spectrum, in this
case in the continuous spectrum at large distances
the interaction potential acquires "knot" points, which are
invariant w.r.t. the choice of gauge. The position of the
"knots" does not depend on the value of the gauge parameter
$d^{(0)}$ and is determined only by the binding energy of the system.

We shall also point out a possibility for a new physical
interpretation for the parameter $\kappa $, which is known not to be
calculated in the original fundamental model, but appears as
reflection of the widely discussed phenomenon of dimensional
transmutation. In ref.[12] this parameter was determined
phenomenologically from the slope of the linearly growing part
of the interaction potential with an account of the
spectroscopy data. In our approach the parameter $\kappa $ being the
quantity of the same measure as the binding energy,
fixes a new scale of distances determining the infrared region.
It is remarkable that the quantity $\kappa $ enters the
interaction potential in the form of a dimensionless ratio
$\beta \equiv (\kappa /\varepsilon )^2$, in this, explicit expressions
(\ref{42}) and (\ref{43}) obviously show that the
quantity $\beta $ characterizes the intensity of the infrared region
influence on the behaviour of the quark--quark forces.

In the present work we pursued the goal to elucidate how the property  of
asymptotic freedom from quantum chromodynamics manifests itself in
terms of a quark--quark interaction quasipotential.

To  achieve  the  goal,  we  have   calculated   the  quark--quark
interaction quasi\-potential applying the single--time
reduction  technique  and  obtained  the  corresponding   explicit
analytical expressions.  In
doing so, the one--loop approximation for the invariant  charge  in
quantum chromodynamics was used.

The analysis of the  resulting  expressions  for  the  interaction
quasipotential has shown  that  here the same scenario is observed
that the transition
from the discrete spectrum region (negative values of the  binding
energy) to the continuous spectrum (region of positive values of  the
binding energy) changes the  pattern in the behavior of quark--quark
forces. In this manner the  nontrivial energy  dependence  of  the  quark
interaction quasipotential manifests itself. The obtained energy
dependence of quark--quark forces, as it has been
repeatedly stressed to be a  consequence  of  a
consistent  consideration  of  the  problem of interaction of
relativistic systems.

There is one more situation described in this work where the
energy  dependence of quark--quark forces shows, and
this one is related with the behavior  of  the  running   coupling
constant in the configuration space at  small  distances.  The
behavior of the running coupling constant proves to be  such  that
when the binding energy is not zero, the  Coulomb  singularity  of
the quark interaction quasipotential at zero is preserved, and it
is only at zero  binding  energy  limit this  Coulomb
singularity is logarithmically  "smoothed  over".  Therefore,  the
known Richardson's  potential  and  its relativized associations
concerning the behavior of quark--quark forces are restored only in
the limit at zero binding energy.
It will of great interest to study the correspondence of the
indicated properties of quark--quark forces and experimental data
on spectroscopy and decays of quarkonium systems.
\begin{center}
\section*{Acknowledgments}
\end{center}
This work was supported in part by the International Science
Foundation. I~am grateful for ISF Grant which enabled to
prepare this report.
\vspace{1cm}




\section*{Appendix A}

Here we present the results of calculating of
integrals which determine the functions $A^{(i)}$ and
$B^{(i)}$, $(i=0,1)$, in the case of
unequal masses of quarks and antiquarks in arbitrary evolution
gauge. The calculation of the integrals yields
\be
A^{(1)}(nM\mid \tilde{p}_1\tilde{p}_2;\tilde{k}_1
\tilde{k}_2)\equiv A^{(1)}(M,P_{\perp }\mid p_{\perp };k_{\perp
})=\label{C.1}\\
\ee
\[
= \frac{(g\kappa )^2}{(2\pi )^3}\cdot \frac{1}{-4(p_{\perp }-k_{\perp
})^2}\times
\nonumber\\
\]
\[
\times [\frac{1}{(\sqrt{m_1^2-p_{1\perp }^2}+
\sqrt{m_2^2-k_{2\perp }^2}+\sqrt{-(p_{\perp }-k_{\perp
})^2}-M)^2}+\nonumber\\
\]
\[
+\frac{1}{(\sqrt{m_2^2-p_{2\perp }^2}+
\sqrt{m_1^2-k_{1\perp }^2}+\sqrt{-(p_{\perp }-k_{\perp
})^2}-M)^2}+\nonumber\\
\]
\[
+\frac{1}{\sqrt{-(p_{\perp }-k_{\perp })^2}
(\sqrt{m_1^2-p_{1\perp }^2}+\sqrt{m_2^2-k_{2\perp }^2}+
\sqrt{-(p_{\perp }-k_{\perp })^2}-M)}+\nonumber\\
\]
\[
+\frac{1}{\sqrt{-(p_{\perp }-k_{\perp })^2}
(\sqrt{m_2^2-p_{2\perp }^2}+\sqrt{m_1^2-k_{1\perp }^2}+
\sqrt{-(p_{\perp }-k_{\perp })^2}-M)}],\nonumber
\]
\be
B^{(1)}(nM\mid \tilde{p}_1\tilde{p}_2;\tilde{k}_1
\tilde{k}_2)\equiv B^{(1)}(M,P_{\perp }\mid p_{\perp };k_{\perp
})=\label{C.2}\\
\ee
\[
= -\frac{(g\kappa )^2}{(2\pi )^3}\cdot \frac{1}
{(2\sqrt{-(p_{\perp }-k_{\perp })^2})^3}\times \nonumber\\
\]
\[
\times \left\{\frac{[\frac{1}{2}(\sqrt{m_1^2-p_{1\perp }^2}+
\sqrt{m_2^2-p_{2\perp }^2}+
\sqrt{m_1^2-k_{1\perp }^2}+\sqrt{m_2^2-k_{2\perp }^2})-M]^2}
{(\sqrt{m_1^2-p_{1\perp }^2}+\sqrt{m_2^2-k_{2\perp }^2}+
\sqrt{-(p_{\perp }-k_{\perp })^2}-M)^3}\right.+ \nonumber\\
\]
\[
+\frac{[\frac{1}{2}(\sqrt{m_1^2-p_{1\perp }^2}+
\sqrt{m_2^2-p_{2\perp }^2}+
\sqrt{m_1^2-k_{1\perp }^2}+\sqrt{m_2^2-k_{2\perp }^2})-M]^2}
{(\sqrt{m_2^2-p_{2\perp }^2}+\sqrt{m_1^2-k_{1\perp }^2}+
\sqrt{-(p_{\perp }-k_{\perp })^2}-M)^3}+ \nonumber\\
\]
\[
+\frac{3(\sqrt{m_1^2-p_{1\perp }^2}-\sqrt{m_2^2-p_{2\perp }^2}-
\sqrt{m_1^2-k_{1\perp }^2}+\sqrt{m_2^2-k_{2\perp }^2})^2}
{2\sqrt{-(p_{\perp }-k_{\perp })^2}}\times \nonumber\\
\]
\[
\times [\frac{1}{2}(\sqrt{m_1^2-p_{1\perp }^2}+
\sqrt{m_2^2-p_{2\perp }^2}+
\sqrt{m_1^2-k_{1\perp }^2}+\sqrt{m_2^2-k_{2\perp }^2})-M]^2\times
\nonumber\\
\]
\[
\times [(\sqrt{m_1^2-p_{1\perp }^2}+\sqrt{m_2^2-k_{2\perp }^2}+
\sqrt{-(p_{\perp }-k_{\perp })^2}-M)\times \nonumber\\
\]
\[
\times (\sqrt{m_2^2-p_{2\perp }^2}+\sqrt{m_1^2-k_{1\perp }^2}+
\sqrt{-(p_{\perp }-k_{\perp })^2}-M)]^{-2}-\nonumber\\
\]
\[
-3[1-\frac{(\sqrt{m_1^2-p_{1\perp }^2}-\sqrt{m_2^2-p_{2\perp }^2}-
\sqrt{m_1^2-k_{1\perp }^2}+\sqrt{m_2^2-k_{2\perp }^2})^2}
{-4(p_{\perp }-k_{\perp })^2}]\times \nonumber\\
\]
\[
\times [\frac{1}{2}(\sqrt{m_1^2-p_{1\perp }^2}+
\sqrt{m_2^2-p_{2\perp }^2}+
\sqrt{m_1^2-k_{1\perp }^2}+\sqrt{m_2^2-k_{2\perp }^2})-M]\times \nonumber\\
\]
\[
\times [(\sqrt{m_1^2-p_{1\perp }^2}+\sqrt{m_2^2-k_{2\perp }^2}+
\sqrt{-(p_{\perp }-k_{\perp })^2}-M)\times \nonumber\\
\]
\[
\times (\sqrt{m_2^2-p_{2\perp }^2}+\sqrt{m_1^2-k_{1\perp }^2}+
\sqrt{-(p_{\perp }-k_{\perp })^2}-M)]^{-1}-\nonumber\\
\]
\[
-\frac{(\sqrt{m_1^2-p_{1\perp }^2}+\sqrt{m_2^2-p_{2\perp }^2}-
\sqrt{m_1^2-k_{1\perp }^2}-\sqrt{m_2^2-k_{2\perp }^2})^2}
{4}\times \nonumber\\
\]
\[
\times \left[\frac{1}{(\sqrt{m_1^2-p_{1\perp }^2}+\sqrt{m_2^2-k_{2\perp
}^2}+
\sqrt{-(p_{\perp }-k_{\perp })^2}-M)^3}\right.+\nonumber\\
\]
\[
+\frac{1}{(\sqrt{m_2^2-p_{2\perp }^2}+\sqrt{m_1^2-k_{1\perp }^2}+
\sqrt{-(p_{\perp }-k_{\perp })^2}-M)^3}+\nonumber\\
\]
\[
+\frac{3}{2\sqrt{-(p_{\perp }-k_{\perp })^2}(\sqrt{m_1^2-p_{1\perp }^2}+
\sqrt{m_2^2-k_{2\perp }^2}+\sqrt{-(p_{\perp }-k_{\perp
})^2}-M)^2}+\nonumber\\
\]
\[
+\frac{3}{2\sqrt{-(p_{\perp }-k_{\perp })^2}(\sqrt{m_2^2-p_{2\perp }^2}+
\sqrt{m_1^2-k_{1\perp }^2}+\sqrt{-(p_{\perp }-k_{\perp
})^2}-M)^2}+\nonumber\\
\]
\[
+\frac{3}{2(\sqrt{-(p_{\perp }-k_{\perp })^2})^2(\sqrt{m_1^2-p_{1\perp
}^2}+
\sqrt{m_2^2-k_{2\perp }^2}+\sqrt{-(p_{\perp }-k_{\perp
})^2}-M)}+\nonumber\\
\]
\[
+\left.\left.\frac{3}{2(\sqrt{-(p_{\perp }-k_{\perp })^2})^2
(\sqrt{m_2^2-p_{2\perp }^2}+
\sqrt{m_1^2-k_{1\perp }^2}+\sqrt{-(p_{\perp }-k_{\perp })^2}-M)}\right]
\right\}.\nonumber
\]
We used here the following notations
\ber
p_{i\perp }=\tilde{p}_i-(n\tilde{p}_i)n,\quad k_{i\perp }=\tilde{k}_i-
(n\tilde{k}_i)n, \quad i=1,2, \nonumber\\
P_{\perp }=p_{1\perp }+p_{2\perp },\quad K_{\perp }=k_{1\perp }+k_{2\perp
},\quad
P_{\perp }=K_{\perp },\label{C.3}\\
p_{\perp }=p_{1\perp },\quad p_{2\perp }=P_{\perp }-p_{\perp },\quad
k_{\perp }=k_{1\perp },\quad
k_{2\perp }=P_{\perp }-k_{\perp }.\nonumber
\eer
\be
A^{(0)}(nM\mid \tilde{p}_1\tilde{p}_2;\tilde{k}_1
\tilde{k}_2)\equiv A^{(0)}(M,P_{\perp }\mid p_{\perp };k_{\perp
})=\label{C.4}\\
\ee
\[
=\frac{g^2}{(2\pi )^3}\cdot \frac{1}
{2\sqrt{-(p_{\perp }-k_{\perp })^2}}\times \nonumber\\
\]
\[
\times \left[\frac{1}{\sqrt{m_1^2-p_{1\perp }^2}+\sqrt{m_2^2-k_{2\perp
}^2}+
\sqrt{-(p_{\perp }-k_{\perp })^2}-M}\right.+\nonumber\\
\]
\[
+\left.\frac{1}{\sqrt{m_2^2-p_{2\perp }^2}+\sqrt{m_1^2-k_{1\perp }^2}+
\sqrt{-(p_{\perp }-k_{\perp })^2}-M}\right],\nonumber\\
\]
\be
B^{(0)}(nM\mid \tilde{p}_1\tilde{p}_2;\tilde{k}_1
\tilde{k}_2)\equiv B^{(0)}(M,P_{\perp }\mid p_{\perp };k_{\perp
})=\label{C.5}\\
\ee
\[
= \frac{-g^2}{(2\pi )^3}\cdot \left\{
[\frac{1}{2}(\sqrt{m_1^2-p_{1\perp }^2}+
\sqrt{m_2^2-p_{2\perp }^2}+\right.\nonumber\\
\]
\[
+\sqrt{m_1^2-k_{1\perp }^2}+\sqrt{m_2^2-k_{2\perp }^2})-M]^2\times
\nonumber\\
\]
\[
\times [(\sqrt{m_1^2-p_{1\perp }^2}+\sqrt{m_2^2-k_{2\perp }^2}+
\sqrt{-(p_{\perp }-k_{\perp })^2}-M)\times \nonumber\\
\]
\[
\times (\sqrt{m_2^2-p_{2\perp }^2}+\sqrt{m_1^2-k_{1\perp }^2}+
\sqrt{-(p_{\perp }-k_{\perp })^2}-M)]^{-2}\times \nonumber\\
\]
\[
\times \frac{(\sqrt{m_1^2-p_{1\perp }^2}-\sqrt{m_2^2-p_{2\perp }^2}-
\sqrt{m_1^2-k_{1\perp }^2}+\sqrt{m_2^2-k_{2\perp }^2})^2}
{-4(p_{\perp }-k_{\perp })^2}]-\nonumber\\
\]
\[
-[1-\frac{(\sqrt{m_1^2-p_{1\perp }^2}-\sqrt{m_2^2-p_{2\perp }^2}-
\sqrt{m_1^2-k_{1\perp }^2}+\sqrt{m_2^2-k_{2\perp }^2})^2}
{-4(p_{\perp }-k_{\perp })^2}]\times \nonumber\\
\]
\[
\times \frac{\frac{1}{2}(\sqrt{m_1^2-p_{1\perp }^2}+
\sqrt{m_2^2-p_{2\perp }^2}+
\sqrt{m_1^2-k_{1\perp }^2}+\sqrt{m_2^2-k_{2\perp }^2})-M}{2\sqrt{-(p_{\perp
}
-k_{\perp })^2}}\times \nonumber\\
\]
\[
\times [(\sqrt{m_1^2-p_{1\perp }^2}+\sqrt{m_2^2-k_{2\perp }^2}+
\sqrt{-(p_{\perp }-k_{\perp })^2}-M)\times \nonumber\\
\]
\[
\times (\sqrt{m_2^2-p_{2\perp }^2}+\sqrt{m_1^2-k_{1\perp }^2}+
\sqrt{-(p_{\perp }-k_{\perp })^2}-M)]^{-1}-\nonumber\\
\]
\[
-\frac{(\sqrt{m_1^2-p_{1\perp }^2}+\sqrt{m_2^2-p_{2\perp }^2}-
\sqrt{m_1^2-k_{1\perp }^2}-\sqrt{m_2^2-k_{2\perp }^2})^2}
{4}\times \nonumber\\
\]
\[
\times \left[\frac{1}{-4(p_{\perp }-k_{\perp })^2(\sqrt{m_1^2-p_{1\perp
}^2}+
\sqrt{m_2^2-k_{2\perp }^2}+\sqrt{-(p_{\perp }-k_{\perp })^2}-M)^2}\right.+
\nonumber\\
\]
\[
+\frac{1}{-4(p_{\perp }-k_{\perp })^2(\sqrt{m_2^2-p_{2\perp }^2}+
\sqrt{m_1^2-k_{1\perp }^2}+\sqrt{-(p_{\perp }-k_{\perp
})^2}-M)^2}+\nonumber\\
\]
\[
+\frac{1}{4(\sqrt{-(p_{\perp }-k_{\perp })^2})^3(\sqrt{m_1^2-p_{1\perp
}^2}+
\sqrt{m_2^2-k_{2\perp }^2}+\sqrt{-(p_{\perp }-k_{\perp
})^2}-M)}+\nonumber\\
\]
\[
+\left.\left.\frac{1}{4(\sqrt{-(p_{\perp }-k_{\perp })^2})^3
(\sqrt{m_2^2-p_{2\perp }^2}+\sqrt{m_1^2-k_{1\perp }^2}+
\sqrt{-(p_{\perp }-k_{\perp })^2}-M)}\right]\right\}\nonumber
\]
Formulae (\ref{C.4}) and (\ref{C.5}) use the same notations as
(\ref{C.3}).

\section*{Appendix B}

In the present Appendix, the integrals   (\ref{73}),   (\ref{79}), 
(\ref{80}) from the main text  are  calculated  for  the  case  of 
different quark masses and in an arbitrary evolution gauge. 

1. We start with integral (\ref{73}) determining  the  $A$ 
function. This integral is written down in the form 
\be
I_A=\int_Cdz f_A(z)\label{A.1}
\ee
with 
\be
f_A(z)=\left(\frac{1}{\delta +z}+\frac{1}{\delta -z}\right)\left(\frac{1}
{\kappa _1-z}+\frac{1}{z-\kappa _2}\right)\frac{1}{ln[(\kappa '_1-z)(z-\kappa '_2)/\Lambda ^2]}.
\label{A.2}
\ee
The quantities $\delta ,\kappa _{1,2},\kappa '_{1,2}$  are  
defined in the main text, and the path of integration $C$ 
is depicted in Fig. 1. 
The same figure shows  the  analytic  structure  of  the  function 
$f_A(z)$: the function $f_A$ has  the  logarithmic  branch  points 
$z=\kappa '_{1,2}$ and the poles at the points $z=(\pm \delta ,\kappa _{1,2},x_{1,2})$, 
where 
\[
x_{1,2}=(np-nk)\pm \sqrt{(\xi -1)\Lambda ^2-(p-k)^2_{\perp }}.
\]
The path of integration $C$  in  expression  (\ref{A.1})  for  the 
function $I_A$ is determined by bypass  rules  for  the 
singularities   given   in   integral   (\ref{73}),   which    are 
unambiguously derived from  the  causal  structure  of  the  local 
quantum  field  theory.  Integral  (\ref{A.1})  is  calculated  by 
closing the path of integration in either the upper, or the  lower 
half--plane. When the path of integration is closed  in  the  upper 
half--plane, there arises an integral due to the discontinuity  of 
the integrand function on the left cut (see Fig. 1). In this case, 
we have that 
\be
\int_Cdz f_A(z)=2\pi i\sum_L Res f_A(z_L)-\int_{-\infty }^{\kappa '_2}dx 
\Delta _Lf_A(x),\label{A.3}
\ee
where $\Delta _Lf_A$ means  the  discontinuity  of  the  function  $f_A$ 
on the left cut 
\[
\Delta _Lf(x)=f(x+i0)-f(x-i0),\quad x<\kappa '_2,
\]
and  the  residues  are  taken  at  the  poles   $z_L=(-\delta ,\kappa _2,x_2)$. 
Calculating the discontinuity of the function $f_A$  on  the  left 
cut and the residues of the same function at the given poles,  for 
the quantity $I_A$ we get 
\ber
I_A=-2\pi i(\kappa _1-\kappa _2)\left[\frac{1}{(\kappa _1+\delta )(\kappa _2+\delta )ln[-(\kappa '_1+\delta )(\kappa '_2
+\delta )/\Lambda ^2]}\right.-\nonumber\\
-\frac{1}{ln\xi }\cdot\frac{2\delta }{(\kappa _1-\kappa _2)(\delta ^2-\kappa _2^2)}-\frac{2\delta \Lambda ^2}
{(x_1-x_2)(\kappa _1-x_2)(x_2-\kappa _2)(\delta ^2-x_2^2)}-\nonumber\\
\left.-\int_{-\infty }^{\kappa '_2}dx\frac{2\delta }{(\kappa _1-x)(x-\kappa _2)(\delta ^2-x^2)[ln^2
\vert (\kappa '_1-x)(x-\kappa '_2)/\Lambda ^2\vert +\pi ^2]}\right].\label{A.4}
\eer
When deriving expression (\ref{A.4}), we did  allow  for  the 
equality 
\be
(\kappa '_1-\kappa _2)(\kappa _2-\kappa '_2)=\xi \Lambda ^2.\label{A.5}
\ee
Had we calculated integral (\ref{A.1})  by  closing  the  path  of 
integration in the lower half--plane, then instead  of  (\ref{A.3}) 
we would have obtained 
\[
\int_Cdz f_A(z)=-2\pi i\sum_R Res f_A(z_R)+
\int_{\kappa '_1}^{\infty }dx\Delta _Rf_A(x)
\]
with $\Delta _Rf_A$ meaning the discontinuity of the function  $f_A$ on  
the right cut 
\[
\Delta _Rf_A(x)=f(x+i0)-f(x-i0),\quad x>\kappa '_1.
\]
The   residues   in   this   case   are   taken   at   the   poles 
$z_R=(\delta ,\kappa _1,x_1)$.  After  the  corresponding   calculations,   we 
obtain the following expression for the quan\-ti\-ty~$I_A$: 
\ber
I_A=-2\pi i(\kappa _1-\kappa _2)\left[\frac{1}{(\delta -\kappa _1)(\delta -\kappa _2)ln[-(\delta -\kappa '_1)(\delta -\kappa '_2
)/\Lambda ^2]}\right.-\nonumber\\
-\frac{1}{ln\xi }\cdot\frac{2\delta }{(\kappa _1-\kappa _2)(\delta ^2-\kappa _1^2)}-\frac{2\delta \Lambda ^2}
{(x_1-x_2)(\kappa _1-x_1)(x_1-\kappa _2)(\delta ^2-x_1^2)}-\nonumber\\
\left.-\int^{\infty }_{\kappa '_1}dx\frac{2\delta }{(\kappa _1-x)(x-\kappa _2)(\delta ^2-x^2)[ln^2
\vert (\kappa '_1-x)(x-\kappa '_2)/\Lambda ^2\vert +\pi ^2]}\right].\label{A.6}
\eer
Here we take into account that 
\be
(\kappa '_1-\kappa _1)(\kappa _1-\kappa '_2)=\xi \Lambda ^2.\label{A.7}
\ee
Expression  (\ref{A.6})  for  $I_A$  is  derived  from  expression 
(\ref{A.4})  through   the   substitution   $\kappa _1\to -\kappa _2,   \kappa _2\to -\kappa _1, 
\kappa '_1\to -\kappa '_2,  \kappa '_2\to -\kappa '_1$.  As   a   consequence,   the   following 
substitution takes place: $x_1\to -x_2,x_2\to -x_1$. As  can  easily  be 
seen, integral (\ref{A.1}) for $I_A$ remains invariant  under  the 
substitution. Therefore, (\ref{A.4})  and  (\ref{A.6})  yield  two 
equivalent representations for the quantity $I_A$. By  taking  the 
half--sum of expressions (\ref{A.4}) and (\ref{A.6}), we can get  a 
symmetric representation for the quantity $I_A$. 

If quark  masses  are  assumed  to  be  equal,  and  a  particular 
evolution gauge (Markov--Yukawa gauge) is chosen, then, remembering 
that in this case 
\[
\kappa _1=-\kappa _2\equiv \kappa _0=\sqrt{-(p-k)^2_{\perp }}\equiv q,
\]
\[
\kappa '_1=-\kappa '_2\equiv \kappa '_0=\sqrt{\xi \Lambda ^2+q^2},
\]
\be
x_1=-x_2\equiv x_0=\sqrt{(\xi -1)\Lambda ^2+q^2},\label{A.8}
\ee
from (\ref{A.4}) or (\ref{A.6}) we obtain
\ber
I_A=-2\pi i\cdot 2\kappa _0\left[\frac{1}{(\delta ^2-\kappa _0^2)
ln[(\kappa '_0\,^2-\delta ^2)/\Lambda ^2]}\right.-\nonumber\\
-\frac{1}{ln\xi }\cdot\frac{\delta }{\kappa _0(\delta ^2-\kappa _0^2)}+\frac{\delta }{(\xi -1)x_0
(\delta ^2-x_0^2)}-\nonumber\\
-\left.\int_{\kappa '_0}^{\infty }dx\frac{2\delta }{(x^2-\delta ^2)(x^2-\kappa _0^2)}\cdot
\frac{1}{ln^2[(x^2-\kappa '_0\,^2)/\Lambda ^2]+\pi ^2}\right].\label{A.9}
\eer
Substituting the  integration  variable  $x^2-q^2=\Lambda ^2y^2$  in  the 
R.H.S. of Eq. (\ref{A.9}) and taking into  account 
expression (\ref{A.8}) for  the  quantities  $\kappa _0,\kappa '_0,x_0$,  from 
(\ref{A.9}) we get representation (\ref{74}) for the function $A$.

2. Write integral (\ref{79}), specifying the  function  $B^{(1)}$, 
in the form analogous to (\ref{A.1})
\be
I_B^{(1)}=\int_Cdz f_B^{(1)}(z),\label{A.10}
\ee
where 
\be
f_B^{(1)}(z)=\left(\frac{1}{\delta +z}+\frac{1}{\delta -z}\right)\left(\frac{1}
{\kappa _1-z}+\frac{1}{z-\kappa _2}\right)^2\frac{x^2}
{ln[(\kappa '_1-z)(z-\kappa '_2)/\Lambda ^2]}.\label{A.11}
\ee
In the $z$ plane, we integrate over the same path $C$ as  depicted 
in Fig. 1. The function $f_B^{(1)}$  has  the  logarithmic  branch 
points  $z=\kappa '_{1,2}$,  simple poles at $z=(\pm \delta ,x_{1,2})$, and the
poles of order 2 at the points $z=\kappa _{1,2}$. As before,  integral 
(\ref{A.10}) is calculated by closing the path of  integration  in 
the upper or lower half--plane, after which  the  residues  at  the 
corresponding  poles   and   discontinuities   of   the   function 
$f_b^{(1)}$  on  the  corresponding  cuts  are  found.  The  final 
result for the quantity $I_B^{(1)}$ is given  in  the  symmetric 
form 
\ber
I_B^{(1)}=2\pi i(\kappa _1-\kappa _2)\left[-\frac{\delta ^2}{(\delta +\kappa _1)^2(\delta +\kappa _2)ln[-(\delta 
+\kappa '_1)(\delta +\kappa '_2)/\Lambda ^2]}\right.-\nonumber\\
-\frac{\delta ^2}{(\delta -\kappa _2)^2(\delta -\kappa _1)ln[-(\delta -\kappa '_1)(\delta -\kappa '_2)/\Lambda ^2]}+\nonumber\\
+\frac{1}{ln\xi }\cdot\frac{2\delta }{(\kappa _1-\kappa _2)^2}\left(\frac{\kappa _1^2}
{\delta ^2-\kappa _1^2}+\frac{\kappa _2^2}{\delta ^2-\kappa _2^2}\right)+\frac{\Lambda ^2}{x_1-x_2}\times 
\nonumber\\
\times \left(\frac{2\delta x_2^2}{(\delta ^2-x_2^2)(\kappa _1-x_2)^2
(x_2-\kappa _2)}+\frac{2\delta x_1^2}{(\delta ^2-x_1^2)(x_1-\kappa _2)^2
(\kappa _1-x_1)}\right)+\nonumber
\eer
\[
+\int_{-\infty }^{\kappa '_2}dx\frac{2\delta x^2}{(\delta ^2-x^2)(\kappa _1-x)^2(x-\kappa _2)
[ln^2\vert (\kappa '_1-x)(x-\kappa '_2)/\Lambda ^2\vert +\pi ^2]}+
\]
\be
+\left.\int^{\infty }_{\kappa '_1}dx\frac{2\delta x^2}{(\delta ^2-x^2)(\kappa _1-x)(x-\kappa _2)^2
[ln^2\vert (\kappa '_1-x)(x-\kappa '_2)/\Lambda ^2\vert +\pi ^2]}\right] \label{A.12}
\ee
Provided  the  quark  masses  are  equal, in the  above--mentioned 
particular evolution gauge  we come to 
\ber
I_B^{(1)}=2\pi i\cdot2\kappa _0\left[-\frac{2\delta ^2}{(\delta +\kappa _0)(\delta ^2-\kappa _0^2)
ln[(\kappa '_0\,^2-\delta ^2)/\Lambda ^2]}\right.+\nonumber\\
+\frac{1}{ln\xi }\cdot\frac{\delta }{\delta ^2-\kappa _0^2}-\frac{2\delta x_0}{(\xi -1)
(\delta ^2-x_0^2)(\kappa _0+x_0)}+\nonumber\\
+\left.\int^{\infty }_{\kappa '_0}dx\frac{4\delta x^2}{(x^2-\delta ^2)(x^2-\kappa _0^2)(x+\kappa _0)}
\cdot\frac{1}{ln^2[(x^2-\kappa '_0\,^2)/\Lambda ^2]+\pi ^2}\right].\label{A.13}
\eer
Representation (\ref{81}) for the function $B^{(1)}$  follows  now 
from equality  (\ref{A.13}) after the integration variable therein 
is replaced  as  $x^2-q^2=\Lambda ^2y^2$  and  the  explicit  expressions  
(\ref{A.8}) for $\kappa _0,x_0$ are taken into account. 

3. Writing integral  (\ref{80}),  which  determines  the  function 
$B^{(2)}$, in the form 
\be
I_B^{(2)}=\int_Cdz f_B^{(2)}(z),\label{A.14}
\ee
with 
\be
f_B^{(2)}(z)=\left(\frac{1}{\delta +z}+\frac{1}{\delta -z}\right)\left(\frac{1}
{\kappa _1-z}+\frac{1}{z-\kappa _2}\right)^2\frac{1}
{ln[(\kappa '_1-z)(z-\kappa '_2)/\Lambda ^2]},
\label{A.15}
\ee
and repeating the above calculations, for the quantity  $I_B^{(2)}$ 
we get that 
\ber
I_B^{(2)}=2\pi i(\kappa _1-\kappa _2)\left[-\frac{1}{(\delta +\kappa _1)^2(\delta +\kappa _2)ln[-(\delta 
+\kappa '_1)(\delta +\kappa '_2)/\Lambda ^2]}\right.-\nonumber\\
-\frac{1}{(\delta -\kappa _2)^2(\delta -\kappa _1)ln[-(\delta -\kappa '_1)(\delta -\kappa '_2)/\Lambda ^2]}+\nonumber\\
+\frac{1}{ln\xi }\cdot\frac{2\delta }{(\kappa _1-\kappa _2)^2}\left(\frac{1}
{\delta ^2-\kappa _1^2}+\frac{1}{\delta ^2-\kappa _2^2}\right)+\frac{2\delta \Lambda ^2}{x_1-x_2}\times 
\nonumber\\
\times \left(\frac{1}{(\delta ^2-x_2^2)(\kappa _1-x_2)^2
(x_2-\kappa _2)}+\frac{1}{(\delta ^2-x_1^2)(x_1-\kappa _2)^2
(\kappa _1-x_1)}\right)+\nonumber
\eer
\[
+\int_{-\infty }^{\kappa '_2}dx\frac{2\delta }{(\delta ^2-x^2)(\kappa _1-x)^2(x-\kappa _2)
[ln^2\vert (\kappa '_1-x)(x-\kappa '_2)/\Lambda ^2\vert +\pi ^2]}+
\]
\be
+\left.\int^{\infty }_{\kappa '_1}dx\frac{2\delta }{(\delta ^2-x^2)(\kappa _1-x)(x-\kappa _2)^2
[ln^2\vert (\kappa '_1-x)(x-\kappa '_2)/\Lambda ^2\vert +\pi ^2]}\right] \label{A.16}
\ee
If in the resulting expression (\ref{A.16}) we put quark masses to 
be equal and  pick  up  Markov--Yukawa evolution  gauge,  then  it 
follows that 
\ber
I_B^{(2)}=2\pi i\cdot2\kappa _0\left[-\frac{2}{(\delta +\kappa _0)(\delta ^2-\kappa _0^2)
ln[(\kappa '_0\,^2-\delta ^2)/\Lambda ^2]}\right.+\nonumber\\
+\frac{1}{ln\xi }\cdot\frac{\delta }{\kappa _0^2(\delta ^2-\kappa _0^2)}-\frac{2\delta }{(\xi -1)
x_0(\delta ^2-x_0^2)(\kappa _0+x_0)}+\nonumber\\
+\left.\int^{\infty }_{\kappa '_0}dx\frac{4\delta }{(x^2-\delta ^2)(x^2-\kappa _0^2)(x+\kappa _0)}
\cdot\frac{1}{ln^2[(x^2-\kappa '_0\,^2)/\Lambda ^2]+\pi ^2}\right].\label{A.17}
\eer
Finally, the above substitution of the integration variable in the 
integral of the R.H.S. of Eq. (\ref{A.17}) and the account of  the 
explicit expressions  (\ref{A.8})  for  the  quantities  $\kappa _0,x_0$ 
bring us to representation (\ref{83}) for the function $B^{(2)}$.

\section*{Appendix C}

When one  goes  over  to  the  configuration  space,  one  has  to 
calculate the integrals of the form 
\be
\tilde F(\varepsilon ,\Lambda ;r)=\int_{0}^{\infty }qdq sin(qr) F(\varepsilon ,\Lambda ;q). \label{B.1}
\ee
Since in the momentum representations the functions we are working 
with are dependent on the dimensionless variables 
\be
F(\varepsilon ,\Lambda ;q)=\frac{1}{\Lambda ^2}f(\varepsilon _\Lambda ;q_\Lambda ),\label{B.2}
\ee
the integral transformation (\ref{B.1}) is also rewrite in terms  of  the 
dimensionless variables 
\be
\tilde f(\varepsilon _\Lambda ;\Lambda r)=\int_{0}^{\infty }q_\Lambda dq_\Lambda  sin(\Lambda rq_\Lambda ) f(\varepsilon _\Lambda ;q_\Lambda ). 
\label{B.3}
\ee

1. The calculation of integral (\ref{92}) for the  function  $A_R$ 
reduces to calculating five integrals of the form (\ref{B.3})  for 
the five functions, corresponding to the five terms in the  R.H.S. 
of equality (\ref{91}). Let us write down these five functions:
\[
f_A^{(1)}(\varepsilon _\Lambda ;q_\Lambda )=\frac{1}{[q_\Lambda ^2-(\varepsilon _\Lambda +i0)^2]ln[1+q_\Lambda ^2-(\varepsilon _\Lambda +i0)^2]},
\]
\[
f_A^{(2)}(\varepsilon _\Lambda ;q_\Lambda )=\frac{1}{q_\Lambda [q_\Lambda ^2-(\varepsilon _\Lambda +i0)^2]},\quad
f_A^{(3)}(\varepsilon _\Lambda ;q_\Lambda )=\frac{1}{q_\Lambda ^3[q_\Lambda ^2-(\varepsilon _\Lambda +i0)^2]},
\]
\[
f_A^{(4)}(\varepsilon _\Lambda ;q_\Lambda )=\frac{1}{q_\Lambda [q_\Lambda ^2-(\varepsilon _\Lambda +i0)^2]^2},
\]
\be
f_A^{(5)}(\varepsilon _\Lambda ;q_\Lambda )=\frac{1}{\sqrt{y^2+q_\Lambda ^2}[y^2+q_\Lambda ^2-
(\varepsilon _\Lambda +i0)^2]}.\label{B.4}
\ee

Integral (\ref{B.3}) of the function $f_A^{(1)}$ may be  rewritten  in 
the form 
\be
\tilde f_A^{(1)}(\varepsilon _\Lambda ;\Lambda r)=\frac{1}{2i}\int_{-\infty }^{\infty }q_\Lambda dq_\Lambda 
\frac{exp(i\Lambda rq_\Lambda )}{[q_\Lambda ^2-(\varepsilon _\Lambda +i0)^2]ln[1+q_\Lambda ^2-(\varepsilon _\Lambda +i0)^2]}.
\label{B.5}
\ee
Now we represent it as a path integral 
\be
\tilde f_A^{(1)}(\varepsilon _\Lambda ;\Lambda r)=\frac{1}{2i}\int_{C}dzg_A^{(1)}(\varepsilon _\Lambda ;\Lambda r,z),
\label{B.6}
\ee
where 
\be
g_A^{(1)}(\varepsilon _\Lambda ;\Lambda r,z)=\frac{zexp(i\Lambda rz)}{(z^2-\varepsilon _\Lambda ^2)ln(1+z^2-\varepsilon _\Lambda ^2)},
\label{B.7}
\ee
and the path of integration is shown in Fig. 2. 

Suppose  that  $0<\varepsilon _\Lambda <1$.  Then  the  analytic  structure  of  the 
function $g_A^{(1)}$ is as in Fig. 2: the function $g_A^{(1)}$ has 
the   logarithmic   branch   points    $z=\pm i\sqrt{1-\varepsilon _\Lambda ^2}$    and 
the poles of oder 2 at the points $z=\pm \varepsilon _\Lambda $. Integral (\ref{B.6}) is 
calculated by  closing  the  path  of  integration  in  the  upper 
half--plane. As a result, we get 
\ber
\tilde f_A^{(1)}(\varepsilon _\Lambda ;\Lambda r)=\frac{\pi }{4}e^{i\Lambda r\varepsilon _\Lambda }(1+\frac{i\Lambda r}{\varepsilon _\Lambda })-
\nonumber\\[2ex]
-\pi \int_{\sqrt{1-\varepsilon _\Lambda ^2}}^{\infty }ydy\frac{exp(-\Lambda ry)}{(y^2+\varepsilon _\Lambda ^2)
[ln^2(y^2+\varepsilon _\Lambda ^2-1)+\pi ^2]}.\label{B.8}
\eer
Were the binding energy is negative: $\varepsilon _\Lambda =-\bar \varepsilon _\Lambda <0, 0<\bar \varepsilon _\Lambda <1$, 
the result would have been 
\[
\tilde f_A^{(1)}(\varepsilon _\Lambda ;\Lambda r)=\frac{\pi }{4}e^{-i\Lambda r\bar \varepsilon _\Lambda }(1-\frac{i\Lambda r}
{\bar \varepsilon _\Lambda })-
\]
\[
-\pi \int_{\sqrt{1-\bar \varepsilon _\Lambda ^2}}^{\infty }ydy\frac{exp(-\Lambda ry)}{(y^2+\bar \varepsilon _\Lambda ^2)
[ln^2(y^2+\bar \varepsilon _\Lambda ^2-1)+\pi ^2]}.
\]
Hence, the result (\ref{B.8}) is true for  the  region  $\vert \varepsilon _\Lambda \vert <1$. 
The transition to the region $\vert \varepsilon _\Lambda \vert >1$ is given  by  the  analytic 
continuation of expression (\ref{B.8}). It is possible to  replace 
the integration variable in (\ref{B.8}): $y^2+\varepsilon _\Lambda ^2=x^2$, bringing 
it thus to the form 
\[
\int_{1}^{\infty }\frac{dy}{y}\frac{exp(-\Lambda r\sqrt{y^2-\varepsilon _\Lambda ^2})}
{ln^2(y^2-1)+\pi ^2},
\]
which is, really, applied in formula (\ref{93}).

The integrals like (\ref{B.3}) for the functions $f_A^{(2,3,4)}$ 
can easily be reduced to those from the Tables. So, here only  the 
results are given.
\[
\tilde f_A^{(2)}(\varepsilon _\Lambda ;\Lambda r)=\frac{1}{\varepsilon _\Lambda }\left[\frac{\pi }{2}e^{i\Lambda r\varepsilon _\Lambda }-
a(\Lambda r\varepsilon _\Lambda )\right],
\]
\[
\tilde f_A^{(3)}(\varepsilon _\Lambda ;\Lambda r)=\frac{1}{\varepsilon _\Lambda ^2}\left\{\frac{1}{\varepsilon _\Lambda }\left[\frac
{\pi }{2}e^{i\Lambda r\varepsilon _\Lambda }-a(\Lambda r\varepsilon _\Lambda )\right]-\frac{\pi }{2}\Lambda r\left[i+\frac{2}{\pi }
b(\mu r)\right]\vert _{\mu \to 0}\right\},
\]
\ber
\lefteqn{\tilde f_A^{(4)}(\varepsilon _\Lambda ;\Lambda r)=}\nonumber\\
&=&\frac{1}{2\varepsilon _\Lambda }\left\{-\frac{1}{\varepsilon _\Lambda ^2}\left[\frac{\pi 
}{2}e^{i\Lambda r\varepsilon _\Lambda }-a(\Lambda r\varepsilon _\Lambda )\right]+\frac{\Lambda r}{\varepsilon _\Lambda }\left[i\frac{\pi }{2}
e^{i\Lambda r\varepsilon _\Lambda }+b(\Lambda r\varepsilon _\Lambda )\right]\right\}\label{B.9}
\eer
with the functions $a(x)$ and $b(x)$  defined  in  the  main  text. 
Expressions  (\ref{B.9}),  which 
were obtained for positive values of the binding energy, hold also 
for its negative values. To verify this, one has to keep  in  mind 
that the functions $a$ and $b$  have  logarithmic  cuts,  and  the 
values of these functions at the cuts are given by  the  relations 
\[
a(e^{\pm i\pi }x)=-a(x)+\pi e^{\mp ix},\quad b(e^{\pm i\pi }x)=b(x)\mp i\pi e^{\mp ix},
\quad x>0.
\]

Integral (\ref{B.3}) for the function $f_A^{(5)}$ is rewritten  in 
the form of 
(\ref{B.5})
\be
\tilde f_A^{(5)}(\varepsilon _\Lambda ;\Lambda r)=\frac{1}{2i}\int_{-\infty }^{\infty }zdz
\frac{exp(i\Lambda rz)}{\sqrt{y^2+z^2}(y^2+z^2-\varepsilon _\Lambda ^2)}.
\label{B.10}
\ee
The analytic structure of the integrand function  in  (\ref{B.10}) 
can be seen in Fig. 3: there  are  root  type  branchings  at  the 
points  $z=\pm iy$ and simple poles at $z=\pm i\sqrt{y^2-\varepsilon _\Lambda ^2}$.  After 
closing the path of integration in the upper half--plane, we find 
that 
\be
\tilde f_A^{(5)}(\varepsilon _\Lambda ;\Lambda r)=\frac{\pi }{2\vert \varepsilon _\Lambda \vert }e^{-\Lambda r\sqrt{y^2-\varepsilon _\Lambda ^2}}-
\int_{y}^{\infty }\frac{dx x exp(-\Lambda rx)}{\sqrt{x^2-y^2}(x^2-y^2+\varepsilon _\Lambda ^2)}.
\label{B.11}
\ee
In  the  obtained  expression  (\ref{B.11}),  one  can  make   the 
substitution of the integration variable: $x^2-y^2=t^2$,  bringing 
it thus to the form 
\[
\int_{0}^{\infty }dt\frac{exp(-\Lambda r\sqrt{t^2+y^2})}{t^2+\varepsilon _\Lambda ^2}.
\]
The  calculations  of  the  latter  with  the  help  of  the 
residue theory technique yield 
\[
\int_{0}^{\infty }dt\frac{exp(-\Lambda r\sqrt{t^2+y^2})}{t^2+
\varepsilon _\Lambda ^2}=\nonumber\\
\]
\[
=\frac{\pi exp(-\Lambda r\sqrt{y^2-\varepsilon _\Lambda ^2})}{2\vert \varepsilon _\Lambda \vert }-\int_{0}^{\infty }
\frac{tdt sin(\Lambda rt)}{\sqrt{t^2+y^2}(t^2+y^2-\varepsilon _\Lambda ^2)}
\]
which, allowing for (\ref{B.11}), brings us back  to  the  initial 
representation (\ref{B.3}) for the function $\tilde f_A^{(5)}$.  That  is 
why we did not use result (\ref{B.11}) in the  bulk of the text. 

2. The calculation of integral (\ref{92}) for the  function  $B_R$ 
also reduces to the calculation of  five  integrals  of  the  form 
(\ref{B.3}) for the five  functions,  corresponding  to  the  five 
terms in the R.H.S. of equality (\ref{95}). For the  reasons  just 
mentioned, we will not integrate the fifth term of (\ref{95})  but 
leave  it  unchanged  in  the   initial   form.   The   functions, 
corresponding to the second, third and fourth terms of (\ref{95}), 
coincide  with  the  functions  $f_A^{(2,3,4)}$.  The  result   of 
integrating such functions has already been given.  So,  the  only 
thing to investigate is integral  (\ref{B.3})  for  the  function, 
corresponding to the first term in the R.H.S. of Eq.(\ref{95})
\be
f_B^{(1)}(\varepsilon _\Lambda ;q_\Lambda )=\frac{1}{q_\Lambda (q_\Lambda -\varepsilon _\Lambda -i0)[q_\Lambda ^2-(\varepsilon _\Lambda +i0)^2]
ln[1+q_\Lambda ^2-(\varepsilon _\Lambda +i0)^2]}.\label{B.12}
\ee
Here from the very beginning one should  distinguish  between  the 
regions of positive and negative values of the binding energy.

 a) The region of positive binding energy values: $\varepsilon _\Lambda >0$. In  this 
case,  integral  (\ref{B.3})  for  the  function  $f_B^{(1)}$   is 
conveniently written in the form 
\ber
\lefteqn{\tilde f_B^{(1)}(\varepsilon _\Lambda ;\Lambda r)=}\nonumber\\
&=&\frac{1}{2i}\int_{-\infty }^{\infty }dq_\Lambda 
\frac{exp(i\Lambda rq_\Lambda )}{[q_\Lambda ^2-(\varepsilon _\Lambda +i0)^2](q_\Lambda -\varepsilon _\Lambda -i0)ln[1+q_\Lambda ^2-(\varepsilon _\Lambda +
i0)^2]}-\nonumber\\[2ex]
&-&\frac{1}{2i}\int_{-\infty }^{0}dq_\Lambda \frac{2\varepsilon _\Lambda exp(i\Lambda rq_\Lambda )}
{[q_\Lambda ^2-(\varepsilon _\Lambda +i0)^2]^2ln[1+q_\Lambda ^2-(\varepsilon _\Lambda +i0)^2]}\equiv I_1-I_2. \label{B.13}
\eer
As in the previous case, we rewrite the integral $I_1$ as  a  path 
integral with the  path  of  integration  shown  in  Fig. 2;  the 
integrand function has a pole of oder 3 at  the  point  $z=\varepsilon _\Lambda $. 
After closing the integration path in  the  upper half--plane  and 
calculating the residue of the integrand function at this pole and 
the discontinuity of the same function on the upper cut, we obtain 
that 
\ber
I_1=-\frac{\pi e^{i\Lambda r\varepsilon _\Lambda }}{8\varepsilon _\Lambda ^2}\left[1+\frac{2}{3}\varepsilon _\Lambda ^2-\frac{3}
{2\varepsilon _\Lambda ^2}+2i\Lambda r(\frac{1}{\varepsilon _\Lambda }-\varepsilon _\Lambda )+(\Lambda r)^2\right]+\nonumber\\[2ex]
+\pi \int_{\sqrt{1-\varepsilon _\Lambda ^2}}^{\infty }dt\frac{exp(-\Lambda rt)}{(t^2+\varepsilon _\Lambda ^2)(t+i\varepsilon _\Lambda )
[ln^2(t^2+\varepsilon _\Lambda ^2-1)+\pi ^2]}.\label{B.14}
\eer
The integral $I_2$ is readily brought to the form 
\be
I_2=-\int_{0}^{\infty }dt\frac{\varepsilon _\Lambda exp(-\Lambda rt)}{(t^2+\varepsilon _\Lambda ^2)^2
[ln\vert 1-\varepsilon _\Lambda ^2-t^2\vert -i\pi \Theta (t^2+\varepsilon _\Lambda ^2-1)]}.\label{B.15}
\ee
Combining  now  (\ref{B.14})  and  (\ref{B.15}),  we  derive   the 
following expression for the function $\tilde f_B^{(1)}$:
\ber
\lefteqn{\tilde f_B^{(1)}(\varepsilon _\Lambda ;\Lambda r)=}\nonumber\\
&=&-\frac{\pi e^{i\Lambda r\varepsilon _\Lambda }}{8\varepsilon _\Lambda ^2}\left[1+\frac{2}{3}\varepsilon _\Lambda ^2-\frac{3}
{2\varepsilon _\Lambda ^2}+2i\Lambda r(\frac{1}{\varepsilon _\Lambda }-\varepsilon _\Lambda )+(\Lambda r)^2\right]+\nonumber\\[2ex]
&+&\int_{0}^{\infty }dt\frac{e^{-\Lambda rt}[\varepsilon _\Lambda ln\vert 1-\varepsilon _\Lambda ^2-t^2\vert +\pi t\Theta (t^2+\varepsilon _\Lambda ^2-1)]}
{(t^2+\varepsilon _\Lambda ^2)^2[ln^2\vert 1-\varepsilon _\Lambda ^2-t^2\vert +\pi ^2\Theta (t^2+\varepsilon _\Lambda ^2-1)]}.\label{B.16}
\eer

b) The region of negative values of the binding energy: $\varepsilon _\Lambda =-\bar 
\varepsilon _\Lambda <0,\bar \varepsilon _\Lambda >0$. Integral (\ref{B.3}) for the  function  $\tilde 
f_B^{(1)}$ in this case is also representable as the difference of 
the two integrals: 
\[
\tilde f_B^{(1)}(\varepsilon _\Lambda ;\Lambda r)=I_1(\varepsilon _\Lambda ;\Lambda r)-I_2(\varepsilon _\Lambda ;\Lambda r),
\]
where 
\ber
\lefteqn{I_1(\varepsilon _\Lambda ;\Lambda r)=}\nonumber\\
&=&\frac{1}{2i}\int_{-\infty }^{\infty }dq_\Lambda 
\frac{exp(i\Lambda rq_\Lambda )}{[q_\Lambda ^2-(\bar \varepsilon _\Lambda -i0)^2](q_\Lambda -\bar \varepsilon _\Lambda +i0)ln[1+q_\Lambda ^2
-(\bar \varepsilon _\Lambda -i0)^2]},\nonumber
\eer
\[
I_2(\varepsilon _\Lambda ;\Lambda r)=\frac{1}{2i}\int_{0}^{\infty }dq_\Lambda \frac{2\bar \varepsilon _\Lambda exp(i\Lambda rq_\Lambda )}
{[q_\Lambda ^2-(\bar \varepsilon _\Lambda -i0)^2]^2ln[1+q_\Lambda ^2-(\bar \varepsilon _\Lambda -i0)^2]}.
\]
We recast the expression for $I_1(\varepsilon _\Lambda ;\Lambda r)$ as the path integral 
\[
I_1(\varepsilon _\Lambda ;\Lambda r)=\frac{1}{2i}\int_{\bar C}dzJ_1(\bar \varepsilon _\Lambda ;\Lambda r,z),
\]
where 
\[
J_1(\bar \varepsilon _\Lambda ;\Lambda r,z)=\frac{exp(i\Lambda rz)}{(z^2-\bar \varepsilon _\Lambda ^2)(z-\bar \varepsilon _\Lambda )
ln(1+z^2-\bar \varepsilon _\Lambda ^2)},
\]
whereas the path of integration $\bar C$ is presented in  Fig. 4. 
The function $J_1(\bar \varepsilon _\Lambda ;\Lambda r,z)$ has a pole of oder 3 at the point 
$z=\bar  \varepsilon _\Lambda $,  a pole of oder 2  at   $z=-\bar   \varepsilon _\Lambda $,   and 
logarithmic branchings at the points   $z=\pm i\sqrt{1-\bar  \varepsilon _\Lambda ^2}$. 
Closing the path  of  integration  in  the  upper half--plane  and 
calculating the residue of the  integrand  function  at  the  pole 
$z=-\bar \varepsilon _\Lambda $ and the discontinuity of this function on the  upper 
cut, for $I_1$ we get that 
\ber
I_1(\varepsilon _\Lambda ;\Lambda r)=\frac{\pi e^{-i\Lambda r\bar \varepsilon _\Lambda }}{8\bar \varepsilon _\Lambda ^2}\left(1-
\frac{3}{2\bar \varepsilon _\Lambda ^2}-\frac{i\Lambda r}{\bar \varepsilon _\Lambda }\right)+\nonumber\\[2ex]
+\pi \int_{\sqrt{1-\bar \varepsilon _\Lambda ^2}}^{\infty }dt\frac{exp(-\Lambda rt)}{(t^2+\bar \varepsilon _\Lambda ^2)
(t+i\bar \varepsilon _\Lambda )[ln^2(t^2+\bar \varepsilon _\Lambda ^2-1)+\pi ^2]}.\label{B.17}
\eer
The integral for $I_2$ is easy reduce to the form 
\be
I_2(\varepsilon _\Lambda ;\Lambda r)=\int_{0}^{\infty }dt\frac{\bar \varepsilon _\Lambda exp(-\Lambda rt)}{(t^2+\bar 
\varepsilon _\Lambda ^2)^2[ln\vert 1-\bar \varepsilon _\Lambda ^2-t^2\vert +i\pi \Theta (t^2+\bar \varepsilon _\Lambda ^2-1)]}.\label{B.18}
\ee
After that, combining  the  two  expressions ---  (\ref{B.17})  and 
(\ref{B.18})--- we  obtain  the  final  result  for  the   function 
$f_B^{(1)}$ to be 
\ber
\tilde f_B^{(1)}(\varepsilon _\Lambda ;\Lambda r)=\frac{\pi e^{-i\Lambda r\bar \varepsilon _\Lambda }}{8\bar \varepsilon _\Lambda ^2}\left(1-
\frac{3}{2\bar \varepsilon _\Lambda ^2}-\frac{i\Lambda r}{\bar \varepsilon _\Lambda }\right)+\nonumber\\[2ex]
+\int_{0}^{\infty }dt\frac{e^{-\Lambda rt}[-\bar \varepsilon _\Lambda ln\vert 1-\bar \varepsilon _\Lambda ^2-t^2\vert +\pi t\Theta (t^2
+\bar \varepsilon _\Lambda ^2-1)]}
{(t^2+\bar \varepsilon _\Lambda ^2)^2[ln^2\vert 1-\bar \varepsilon _\Lambda ^2-t^2\vert +\pi ^2\Theta (t^2+\bar \varepsilon _\Lambda ^2-1)]}
\label{B.19}
\eer

\newpage
\vspace*{-3cm}
\begin{picture}(360,150)(10,10)
\put(0,70){\line(360,0){360}}
\put(200,20){\line(0,110){110}}
\put(170,70){\circle*{4}}
\put(150,70){\circle*{4}}
\put(120,70){\circle*{4}}
\put(90,70.7){\line(-90,0){90}}
\put(90,69.3){\line(-90,0){90}}
\put(230,70){\circle*{4}}
\put(250,70){\circle*{4}}
\put(270,70){\circle*{4}}
\put(290,70.7){\line(130,0){70}}
\put(290,69.3){\line(130,0){70}}
\put(170,70){\oval(8,8)[b]}
\put(150,70){\oval(8,8)[b]}
\put(120,70){\oval(8,8)[b]}
\put(90,70){\oval(5,5)[br]}
\put(90,67){\line(-90,0){90}}
\put(230,70){\oval(8,8)[t]}
\put(250,70){\oval(8,8)[t]}
\put(270,70){\oval(8,8)[t]}
\put(290,70){\oval(5,5)[tl]}
\put(290,73){\line(130,0){70}}
\put(180,0){Fig. 1}
\put(300,120){\circle{15}}
\put(300,120){\makebox(0,0){z}}
\put(300,75){$C$}
\put(330,75){$-i\pi$}
\put(340,58){$i\pi$}
\put(225,55){$\delta$}
\put(245,55){$\kappa_1$}
\put(285,55){$\kappa'_1$}
\put(265,55){$x_1$}
\put(162,75){$-\delta$}
\put(145,75){$\kappa_2$}
\put(115,75){$x_2$}
\put(85,75){$\kappa'_2$}
\end{picture}

\begin{picture}(360,150)(10,10)
\put(0,70){\line(360,0){360}}
\put(200,20){\line(0,110){110}}
\put(150,70){\circle*{4}}
\put(200.7,100){\line(0,30){30}}
\put(199.3,100){\line(0,30){30}}
\put(200.7,40){\line(0,-20){20}}
\put(199.3,40){\line(0,-20){20}}
\put(250,70){\circle*{4}}
\put(150,70){\oval(8,8)[t]}
\put(250,70){\oval(8,8)[b]}
\put(180,0){Fig. 2}
\put(300,120){\circle{15}}
\put(300,120){\makebox(0,0){z}}
\put(172,120){$-i\pi$}
\put(205,120){$i\pi$}
\put(135,60){$-\varepsilon_\lambda$}
\put(245,75){$\varepsilon_\lambda$}
\put(205,95){$i\sqrt{1-\varepsilon^2_\lambda}$}
\put(205,35){$-i\sqrt{1-\varepsilon^2_\lambda}$}
\end{picture}

\begin{picture}(360,150)(10,10)
\put(0,70){\line(360,0){360}}
\put(200,20){\line(0,110){110}}
\put(200.7,100){\line(0,30){30}}
\put(199.3,100){\line(0,30){30}}
\put(200.7,40){\line(0,-20){20}}
\put(199.3,40){\line(0,-20){20}}
\put(180,0){Fig. 3}
\put(300,120){\circle{15}}
\put(300,120){\makebox(0,0){z}}
\put(172,120){$-i\frac{\pi}{2}$}
\put(205,120){$i\frac{\pi}{2}$}
\put(205,75){$i\sqrt{y^2-\varepsilon^2_\lambda}$}
\put(205,55){$-i\sqrt{y^2-\varepsilon^2_\lambda}$}
\put(205,95){$iy$}
\put(205,35){$-iy$}
\put(200,80){\circle*{3}}
\put(200,60){\circle*{3}}
\end{picture}

\begin{picture}(360,150)(10,10)
\put(0,70){\line(360,0){360}}
\put(200,20){\line(0,110){110}}
\put(150,70){\circle*{4}}
\put(200.7,100){\line(0,30){30}}
\put(199.3,100){\line(0,30){30}}
\put(200.7,40){\line(0,-20){20}}
\put(199.3,40){\line(0,-20){20}}
\put(250,70){\circle*{4}}
\put(150,70){\oval(8,8)[b]}
\put(250,70){\oval(8,8)[t]}
\put(180,0){Fig. 4}
\put(300,120){\circle{15}}
\put(300,120){\makebox(0,0){z}}
\put(172,120){$-i\pi$}
\put(205,120){$i\pi$}
\put(140,75){$-\bar\varepsilon_\lambda$}
\put(245,56){$\bar\varepsilon_\lambda$}
\put(205,95){$i\sqrt{1-\bar\varepsilon^2_\lambda}$}
\put(205,35){$-i\sqrt{1-\bar\varepsilon^2_\lambda}$}
\put(300,72){$\bar C$}
\end{picture}


\begin{thebibliography}{**}
\bibitem{1}
Collection "Shift of atomic electron levels". M. Foreign Lit. 1950.
Collection "Recent development of quantum electrodynamics". M. Foreign
Lit. 1954.
\bibitem{2}
D. E. Browm, A. D. Jackson. The nucleon--nucleon interaction.
M.: Atomizdat. 1979.
\bibitem{3}
A. A. Bykov, I. M. Dremin, A. V. Leonidov. Usp.Fiz.Nauk. 143 (1984) 3--32.
\bibitem{4}
C. Quigg, J. L. Rosner. Phys.Rep. 56 (1979) 167.
\bibitem{5}
K. Wilson. Phys.Rep. D10 (1974) 2445--2459.
\bibitem{6}
Yu. M. Makeenko. Usp.Fiz.Nauk. 143 (1984) 161--2212.
\bibitem{7}
M. Fukugita. Lattice Quantum Chromodynamics with Dy\-na\-mi\-cal Quarks.
Preprint RIEP--729, December, 1987. A. Ukawa. Lattice QCD Simulations
beyond the Quenched Approximation. Preprint CERN--TH--5245/88, 1988.
\bibitem{8}
A. A. Arkhipov. Sov. J. TMF. 83 (1990) 247--267.
\bibitem{9}
N. N. Bogoljubov, D. V. Shirkov. Introduction  to  the  Theory  of
quantized Fields. M.: Nauka, 1984.
\bibitem{10}
A. A. Arkhipov. Sov. J. TMF. 83 (1990) 358--373.
\bibitem{11}
A. A. Arkhipov. Int. J. Mod. Phys.A. 7 (1992) 683--708.
\bibitem{12}
B. A. Arbuzov. Sov. J. Part. and Nucl., 19 (1988) 5--50.
\bibitem{13}
E. Yanke, F. Emde, F. Lesh. Special Functions. M.: Nauka, 1984.
\bibitem{14}
D. I. Gross, F. Wilczek. Phys.  Rev.  Lett. 30 (1973) 1343--1346.
\bibitem{15}
H. D. Politzer. Phys. Rev. Lett. 30 (1973) 1346--1349.
\bibitem{16}
J. L. Richardson. Phys. Lett. 82B (1979) 272--274.
\bibitem{17}
A. A. Arkhipov. Sov. J. TMF. 74 (1988) 69--81.
\bibitem{18}
B. A. Arbuzov, E. E. Boos, V. I. Savrin, S. A. Shichanin. Sov. J.
Pisma JETP. 50 (1989) 236--238; Mod. Phys. Lett.A. 5 (1990) 1441--1449.
\end{thebibliography}
\end{document}